\documentclass[manuscript]{aastex}
\usepackage{natbib}
\usepackage{amsmath}
\begin{document}
\newcommand{\crc}{cells$/r_c$}
\newcommand{\rr}[1]{$R_{#1}$}

\shorttitle{Ambient Dynamics in Bonnor-Ebert Collapse}
\shortauthors{Kaminski et al.}

\title{On the Role of Ambient Environments in the Collapse of Bonnor-Ebert Spheres}

\author{Erica Kaminski\altaffilmark{1}, Adam Frank\altaffilmark{1}, Jonathan Carroll\altaffilmark{1}, Phil Myers\altaffilmark{2}}
\altaffiltext{1}{Department of Physics and Astronomy, University of Rochester, 206 Bausch and Lomb Hall, P.O. Box 270171, Rochester, NY 14627-0171 \\Email contact: erica@pas.rochester.edu}

\altaffiltext{2}{Harvard-Smithsonian Center for Astrophysics, 60 Garden Street, Cambridge, MA 02138, USA \\Email contact: pmyers@cfa.harvard.edu}

\begin{abstract}
We consider the interaction between a marginally stable Bonnor-Ebert (BE) sphere and the surrounding ambient medium. In particular, we explore how the infall from an evolving ambient medium can trigger collapse of the sphere using 3-D adaptive mesh refinement simulations. We find the resulting collapse dynamics to vary considerably with ambient density. In the highest ambient density cases, infalling material drives a strong compression wave into the cloud. It is the propagation of this wave through the cloud interior that triggers the subsequent collapse. For lower ambient densities, we find the main trigger of collapse to be a quasistatic adjustment of the BE sphere to gravitational settling of the ambient gas. In all cases, we find that the classic ''outside-in'' collapse mode for super-critical BE spheres is recovered before a protostar (i.e. sink particle) forms. Our work supports scenarios in which BE dynamics naturally begins with either a compression-wave or infall dominated phase, and only later assumes the usual outside-in collapse behavior. 
\end{abstract}

\section{Introduction}

''Bonnor-Ebert''(BE) spheres \citep{bonnor1956, ebert1955} have long been a favorite candidate for initial conditions in studies of star formation \citep{hunter1977, foster1994, banerjee2004}. These non-singular, isothermal, hydrostatic solutions to the the self-gravitating fluid equations are constructed to be in pressure equilibrium with an ambient environment making them an effective ''toy-model'' for investigations of protostellar collapse. In addition, even though BE spheres represent highly simplified solutions to these equations, observations have shown some quasi-spherical clumps to match the hydrostatic equilibrium (HSE) of BE solutions ({\it i.e.} the Bok Globule B68 \citep{alves2001}). 

In terms of theoretical studies, a particularly useful property of BE spheres is their well-established stability criterion. Isothermal, hydrostatic spheres with density contrasts between cloud core and cloud edge below 14.5 are stable against gravitational contraction (first shown by \cite{bonnor1956, ebert1955}). This simple stability condition can be cast in terms of a critical mass ($M_{BE}$), similar to the Jeans mass ($M_J$) but with a slightly different numerical coefficient (e.g. for the critical BE sphere, $M_{BE}$ $\backsim$ 2 $M_J$ of a uniform, isothermal sphere of equal temperature and external pressure). Such a critical mass provides an easy interpretation of the relative importance of thermal and gravitational energies within the sphere, or assumed ''pre-protostellar core'' (a phrase coined by \cite{ward-thompson2007}).  Once this stability condition is violated, collapse is triggered.

While the initial properties of isothermal spheres are easily derived analytically, deriving analytical models for their collapse has not been trivial. Because of this, it has been part of an on-going debate about the true nature of the mechanisms involved in isolated star formation.  Modern studies of the detailed dynamics of protostellar cloud collapse date back to work by Larson and Penston (LP) who explored similarity solutions for uniform initial density clouds (\citep{larson1969, penston1969}). Later, Shu presented the ''expansion wave'' similarity solution for singular ($\rho \propto$ $~r^{-2}$) isothermal spheres (SIS)  \citep{shu1977}. The Shu solution described the collapse of the infinitely centrally condensed SIS, whose center immediately collapses into a constant-accreting point mass. This type of core formation results in an {\it inside-out} collapse; adjacent shells collapse one after the next in an outward propagating expansion wave of lost pressure support. Shu and collaborators \citep{shu1987, adams1987} suggested this inside-out collapse would occur for {\it any} hydrostatic, isothermal, spherically-symmetric distribution, including Bonnor-Ebert spheres. This was speculated to occur through a subsonic approach from the flat-topped density distribution of the BE sphere to the r$^{-2}$ density law characteristic of the SIS. After a sufficient perturbation, it was assumed inside-out collapse would be initiated.

Simulations of isolated, unstable BE spheres, however, showed very different behavior than an outward-propagating rarefaction \citep{hunter1977, foster1993, foster1994,ogino1999, keto2010, banerjee2004}. Unlike the SIS, BE spheres collapse from the {\it outside}-in.  This term 'outside-in collapse' describes an inward directed flow that begins in the outer regions of the BE sphere. The density profile that develops shows the flat-topped inner region to increase in density, while shrinking in radial extent. Beyond the flat inner core region, a $r^{-2}$ density envelope develops. The corresponding radial velocity profiles shows the flow beginning in the outer radii of the sphere to move inward with time, peaking with a Mach number close to 2. At the center of the sphere, the velocity always turns back to zero. Eventually, a central accreting point mass forms, with an accretion rate that tends to decrease with time. Simulations have shown this to be the general progression of BE collapse, irrespective of perturbation type, degree of central concentration, and outer boundary conditions, and has been described in great detail in many papers \citep{hunter1977, foster1993, banerjee2004}.

The complexity of dynamics shown in simulations has made it challenging to reconcile the BE sphere collapse with the early similarity solutions. In \cite{whitworth1985} (from here on referred to as WS) a general set of similarity solutions for the isothermal sphere were derived, with the Shu and LP solutions appearing at varying limits. The WS solutions describe the collapse as either being triggered or enhanced by an externally driven, inward propagating compression wave, depending on the degree of  initial stability of the sphere. At early times the sphere is either in a static equilibrium state or condensing because it is initially far out of equilibrium (e.g. uniform sphere). The head of the compression wave represents a region of enhanced density that propagates deeper into the sphere, and its strength depends on the degree of external pressure at the sphere's surface. Once the compression wave converges on the center, a central accreting object forms.  

A number of authors have since studied the BE collapse problem in the WS compression wave framework. For instance, \cite{hennebelle2003} looked at the effect of differentially increasing the external pressure on marginally stable BE spheres. The rate at which they increased the external pressure varied in their runs, which allowed them to infer a general trend: an inward propagating compression wave was induced at the BE surface that later converged on the center and resulted in a centrally accreting point mass. The strength of the compression wave varied depending on the magnitude and duration of the pressure increase, in line with the WS model. \cite{anathpindika2013} also studied the collapse of the BE sphere in terms of a compression-wave scenario.  They triggered collapse by reducing the thermal energy inside the sphere's interior, thus triggering a compression wave that in some cases amplified condensation and led to collapse, while in other cases only triggered oscillations. 
  
It is noteworthy that despite the great deal of attention paid to BE spheres, numerical studies of their basic collapse properties have placed them in discontinuous, low-density ambient media and then perturbed them into collapse. This implies an isolation of the collapse from the ambient environment. From this, it is hard to disentangle the effect of the ambient medium, which has been included in the WS framework. Further, the effect of density mismatch between the sphere's outer edge and its encompassing medium is also poorly understood. For instance, \cite{myers2008} took care in modeling the transition in density from sphere to ambient gas. In that study, where the infall and dispersal of ambient gas onto BE spheres was calculated, Myers underscored the importance in considering this density transition. It thus seems an interesting and previously unexplored topic to {\it not} isolate the BE sphere in studying its collapse, and instead consider the  role the {\it ambient medium} plays in the evolution of the BE sphere. To this end, we performed a series of numerical experiments that placed a marginally stable BE sphere in the center of mass position of various concentric, homogeneous background media of different densities. We then explored the system's natural response with no added perturbations. 

This paper is organized as follows. In section 2 we describe our methods and numerical models.  In section 3 we present results of our simulations which are then discussed in more detail in section 4.  Section 5 presents our conclusions.  We also provide 2 appendices which detail our treatment and tests of self-gravity in Astrobear.

\section{Methods}\label{methods}
\subsection{Bonnor-Ebert sphere definitions}\label{BEdefs}

\begin{figure}[htbp]
\centering
\epsscale{.60}\plotone{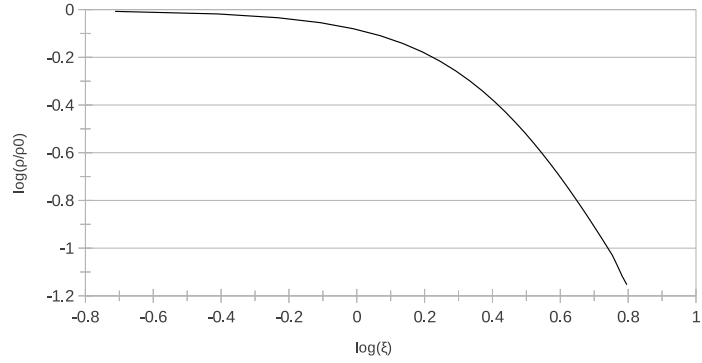}
\caption{{The density profile of a critical Bonnor-Ebert sphere as a function of $\xi$ in log-log space. The y-axis is in scaled units, normalized to the initial central density $\rho_{0}$ of the BE sphere. Given the scaled nature of this curve, it represents a family of solutions, each BE sphere given by a different $\rho_{0}$ and truncation radius.}}
\label{fig_BE}
\end{figure}

All BE spheres show a flat-topped density profile near their centers with outer densities that decline monotonically with radius (Fig.~\ref{fig_BE}). The BE sphere, defined by its truncation radius, $R_{BE}$, and confining pressure, $P_{ext}(R_{BE})$, is thermally or gravitationally dominated depending on the value of its non-dimensional radius,

\begin{equation}\label{xi}
\xi = (\frac{4 \pi G \rho_{0}}{C_{s}^2})^{1/2} r
\end{equation}

\noindent where G is the gravitational constant, $\rho_0$ is the initial central density, $C_s$ is the isothermal sound speed, and r is the dimensional radius. The sphere is semi-stable, i.e. on the cusp of its stability curve, when its $\xi\approx6.5$. At this point, with no added perturbation, the sphere may oscillate gently about its equilibrium. Likewise, collapse can ensue in this state with the application of small perturbations. The critical radius and external pressure of such a marginally stable BE sphere of mass $M_{BE}$ are \citep{spitzer1968}:

\begin{equation}\label{Pcrit}
r_{crit} = 0.41 \frac{G M_{BE}}{C_{s}^2}
\end{equation}

\noindent and

\begin{equation}\label{Rcrit}
P_{crit} = 1.40 \frac{C_{s}^8}{G^3 M_{BE}^2}.
\end{equation}

To initialize our simulations we used an approximate analytical solution of the Lane-Emden equation. This expression yields a nearly perfect solution, with a relative error of less than 1$\%$ compared to the numerical solution \citep{liu1996}. The required inputs to the code were $r$, $\xi$, and $\rho_0$, which output the density profile $\rho(r)$ \footnotemark[3]. Given isothermality, this yielded the pressure profile $P( r )$ as well.

\footnotetext[3]{documentation: astrobear.pas.rochester.edu/trac/wiki/u/BonnorEbertModule}

\subsection{Model}
We modeled the collapse of the marginally stable BE sphere ($\xi = 6.5$) using Astrobear2.0, a highly parallelized, multidimensional, adaptive mesh refinement (AMR) code that solves the equations of hydrodynamics and magnetohydrodynamics on a Cartesian grid \citep{andy2009,carroll2013}. Astrobear2.0 has a library of multiphysics tools, including heat conduction, resistivity, radiative cooling, and self-gravity. In the present work, Astrobear solved the Euler equations with self-gravity. The Poisson solver for self-gravity uses HYPRE \citep{falgout2002} \footnotemark[4], a software package that solves linear systems on massively parallelized systems. Our methods for self-gravity are described in Appendices (\ref{poissonsolver}-\ref{testingpoissonsolver}) . 

\footnotetext[4]{documentation: https://computation.llnl.gov/casc/hypre/software.html}

The problem domain was a cube of side length $L$, with the BE sphere at the center. We initialized the fluid variables the same as in \cite{banerjee2004}, hereafter 'BPH04', to verify Astrobear could reproduce their results. Those authors chose parameters to match potential sites of massive star formation. Thus, our BE sphere has an initial radius of $R_{BE}$ $\approx 1.62 ~pc$, central density $n_0 = 2004 ~cm^{-3}$, temperature $T = 20 ~K$, mass $M_{BE} = 151 ~M_{\sun}$, and isothermal sound speed $C_s \approx 0.4~ km/s$. It is initialized to be in pressure equilibrium with the ambient material. The gas in the box was assumed to be atomic hydrogen ($\mu = 1$). We used an ideal equation of state, with an adiabatic index very close to 1 ($\gamma = 1.0001$), but not exactly 1, to avoid infinities in the routine. This effectively gives the gas a nearly infinite heat capacity, which allows the gas to maintain a nearly constant temperature when compressed. Note, that while the evolution proceeds nearly isothermally, the temperature of the ambient gas would be different than the BE sphere should there be an initial density mismatch at their boundary, given pressure equilibrium. 

For this study we placed the BE sphere at the origin of a spherically symmetric mass distribution inside of the simulation box. This is a special case and models an embedded core (the BE sphere) that is located at the center of mass position of an encompassing, concentric cloud complex. If the parent cloud is gravitationally unstable, its collapse path will be toward the center of the box, coinciding with the position of the BE sphere. This implies a maximally disruptive system -- collapse of the parent cloud will maximally disrupt the BE sphere, and thus this setup represents an idealization of the process we are studying. 

\begin{figure}[htbp]
\centering
\epsscale{0.45}
\plotone{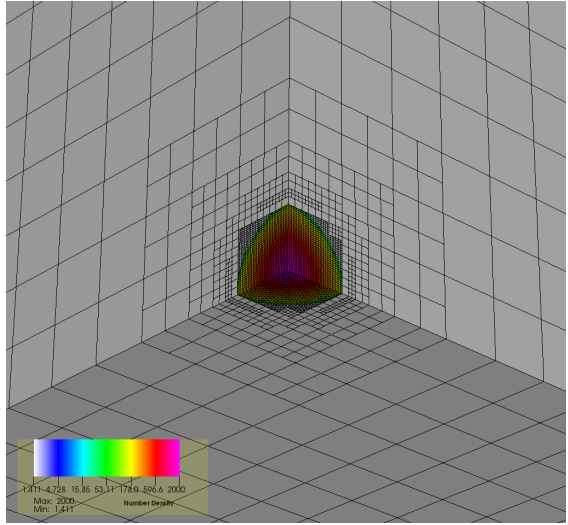}
\caption{{Schematic of the mesh with an octant of the Bonnor-Ebert sphere located at (0,0,0). While all simulations were initialized with 5 levels of refinement to achieve $\backsim$34 cells per initial clump radius, only the first 3 levels are plotted here for clarity. The mesh was dynamic in that additional levels of refinement were added as needed (see text). The color-bar shows variation in $n ~(cm^{-3})$. Note that the maximum value in this plot is lower than the reported 2004 $cm^{-3}$, given the data from the finest cells are not plotted.}}
\label{Mesh}
\end{figure}

Given the symmetry of the problem, we only simulated an octant of the domain with the mesh extending from $(0-25)~pc$ in $x-, ~y-, $ and $z-$ directions and the BE sphere centered at (0,0,0) (currently Astrobear is not configured in spherical coordinates: Fig. \ref{Mesh}). Note that the AMR allows us to place the boundaries of the domain far from the edge of the BE sphere with $L = 30~ R_{BE}$, allowing any potential boundary effects to be mitigated. Boundary conditions on the outward-facing sides of the box were extrapolating with velocity set to 0, preventing gas from leaving the box. On the faces that sliced through the BE sphere, the boundary conditions were reflecting. The coarse grid was made up of $16^3$ cells, initialized with 5 levels of AMR within the BE sphere. Each level of AMR increased the effective number of cells in each direction by a factor of 2. Thus, at the start of the simulation the effective resolution was $\triangle x_{min} \approx 0.05$ pc, corresponding to $\approx$ 34 cells/$R_{BE}$. Two additional levels of refinement were triggered during the BE sphere collapse when the ''Truelove condition'' \citep{truelove1997},

\begin{equation}\label{Truelove}
\triangle x > \lambda_{J} /4
\end{equation}

\noindent was violated, where $\lambda_{J}$ is the local Jeans length associated with each cell center, and $\triangle x$ is the width of a cell on any given level. Essentially this is an AMR criterion that maintains adequate resolution of the Jeans length. Once the maximum refinement is reached, sink particles are generated based on conditions described in \cite{fedderrath2010}. The simulations were halted after the formation of a sink.

\subsection{Simulations}

We performed a set of 9 simulations, listed in Table~\ref{tab_runs}. In each run the marginally stable ($\xi = 6.5$) BE sphere was initialized to be in pressure equilibrium with an ambient medium of a different uniform density. To test the stability of the BE sphere, we chose an ambient density of $\rho_{amb}~(r > R_{BE}) = ~0.01\rho~(R_{BE})$, where again $R_{BE}$ is the truncation radius of the BE sphere. This we named the ''Sparse'' ambient case. No perturbation was applied to the sphere and we ran the simulation out for 5 crossing times $t_{sc}$ of the sphere, 

\begin{equation}
\label{eqn_tsc}
t_{sc} = R_{BE} / C_{s} \approx~ 3.8~ Myr.
\end{equation}

\noindent Next we added a 10$\%$ density enhancement to every cell within the simulation box for the Sparse ambient case. This was the same form of perturbation BPH04 and others have used to initiate outside-in collapse. We designate this the ''Classic'' case. We then carried out a series of runs that placed the sphere in ambient environments of varying density. Note that no perturbations were applied in any of these runs; it was only the effect of the ambient medium on the BE sphere that drove the subsequent dynamics. The ambient density in this series ran from  $\rho_{amb} = \rho~(R_{BE})$ (''Matched'' case) to $\rho_{amb} = 1/50 ~\rho~(R_{BE})$ (see Table \ref{tab_runs}).

\section{Results}\label{results}
\subsection{A Test of Stability - Sparse Ambient Medium}\label{Stability}

\begin{figure}[htbp]
\centering
\epsscale{1}
\plottwo{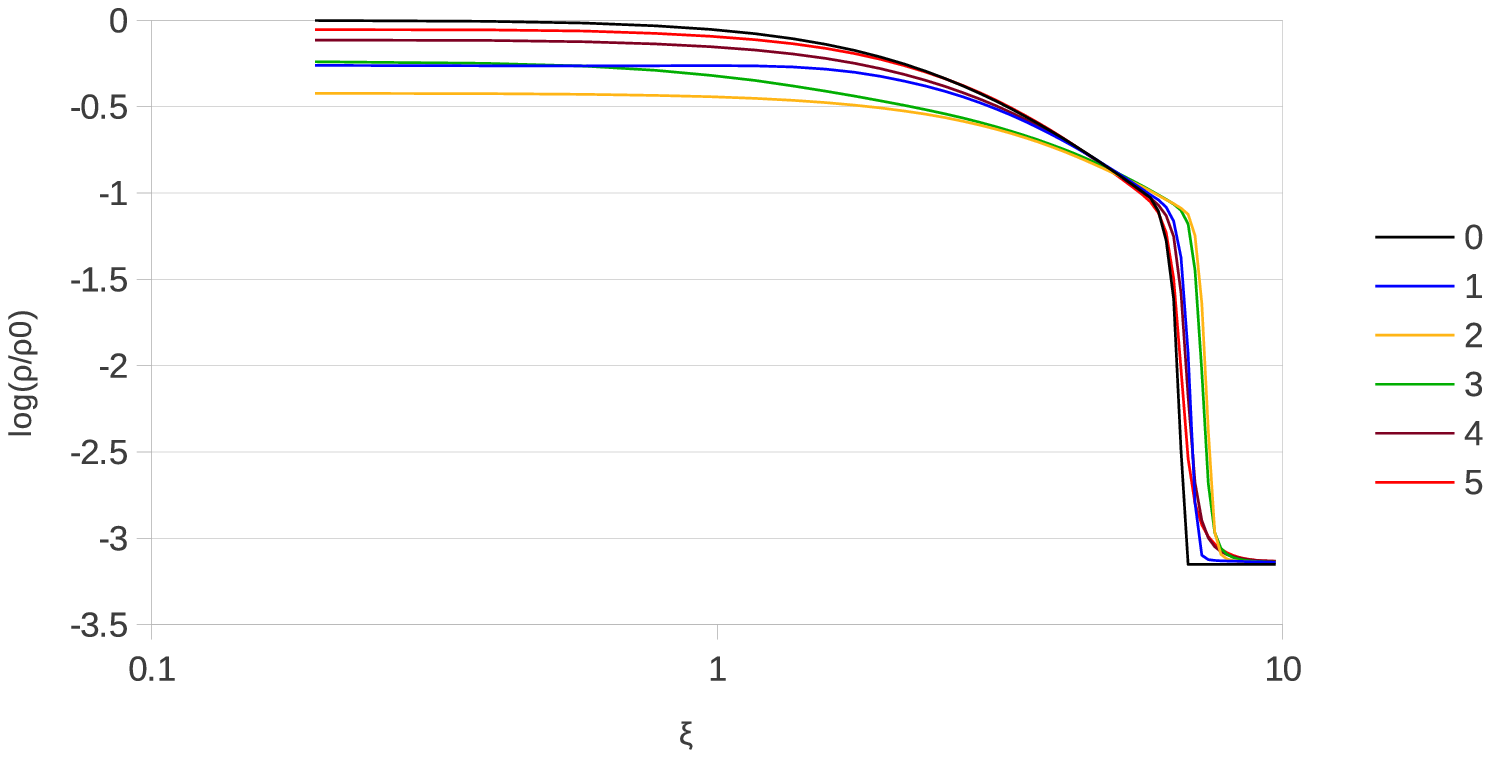}{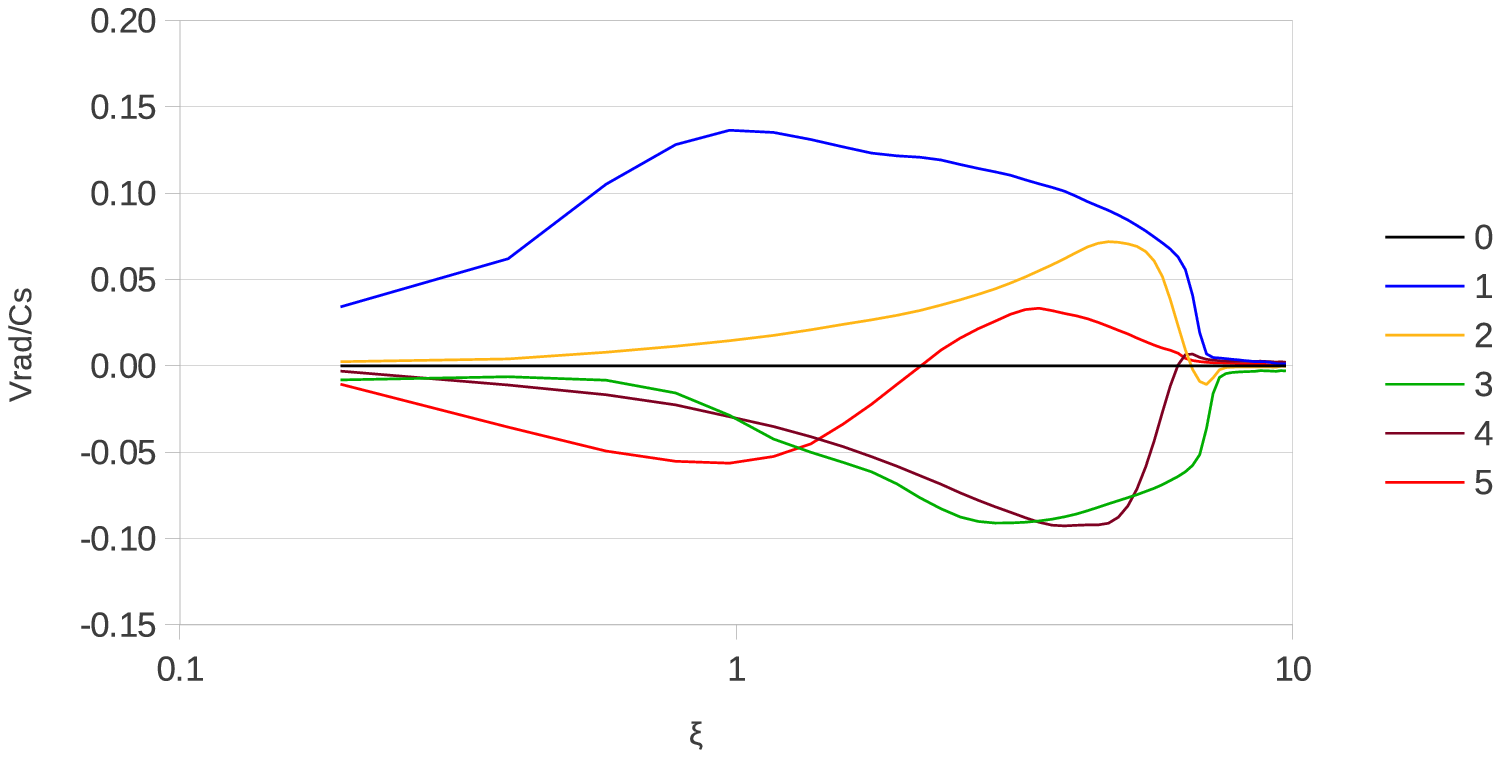}
\caption{{Density and radial velocity profiles for the Sparse case, $\rho_{amb}=\rho(R_{BE})/100$, over time. The legend is time in units of the sphere's sound-crossing time, t/t$_{sc}$. The sphere's initial outer boundary is at $\xi = 6.5$ on the x-axis and can be traced by the sharp discontinuous jump in density that occurs at the BE sphere-ambient boundary. Density and radial velocity are averaged over angle in these and all subsequent plots.  As can be seen on the left, the sphere oscillated to lower $\rho(r)$, but returned close to the equilibrium profile by t/$t_{sc}$ = 5. The right-hand plot shows small subsonic radial motions throughout the sphere during this time period. As usual, negative velocities indicate inward motions.}}
\label{light_case}
\end{figure}

As discussed in the last section, in this run the BE sphere was placed in a hot (2000 K), low density (1.4 $cm~^{-3}$) ambient medium.  The simulation was run for five sound crossing times of the sphere ($\approx$ 19 Myr, eqn. \ref{eqn_tsc}).  Figure~\ref{light_case} shows the density and radial velocity profiles within $r=1.5~R_{BE}$ during the simulation. For all density and radial velocity curves shown in this paper, the data have been averaged over angle to minimize any noise induced by the non-spherical mesh.  Fluid motions remained subsonic and represented sound waves traveling from edge to center as the BE sphere adjusted to the initial setup. The sphere oscillated around its initialized equilibrium values with no indication of either collapse or unstable expansion being triggered. The largest radial excursion in the sphere's radius represented only $10\%$ of the initial value of $R_{BE}$.  These results confirm the code's ability to recover appropriate, analytical solutions for self-gravitating fluids.

\subsection{Classic Collapse}\label{classic}

As a second test of our code, we  reproduced the results of other studies of  BE sphere collapse (namely, BPH04). Holding other conditions equal to the Sparse case, we added a 10$\%$ density increase to the entire simulation box. This increased the mass within the sphere above that which can be supported by thermal pressure alone, as dictated by the Lane-Emden equation ($E_{th} < E_g$). 

\begin{figure}[htbp]
\centering
\plottwo{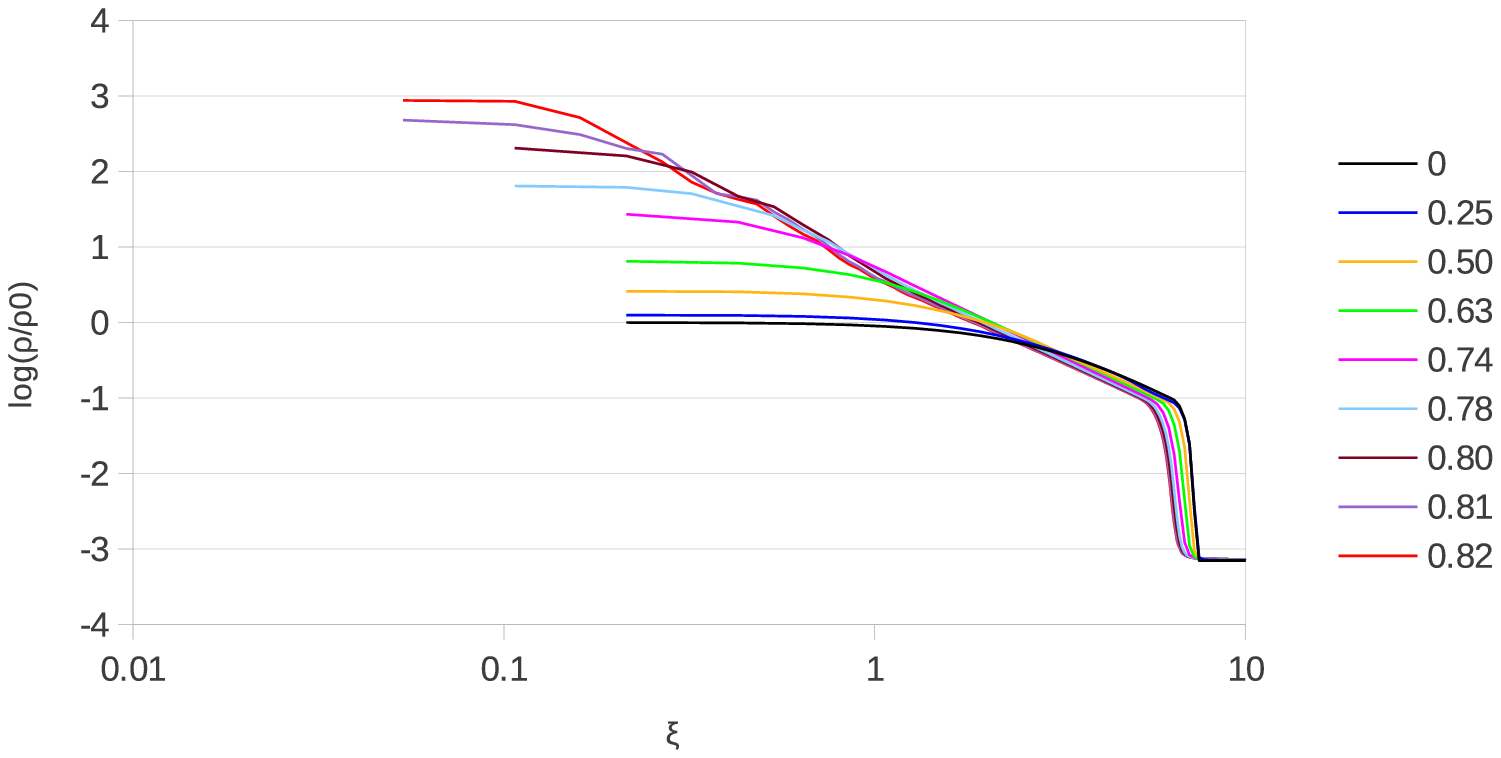}{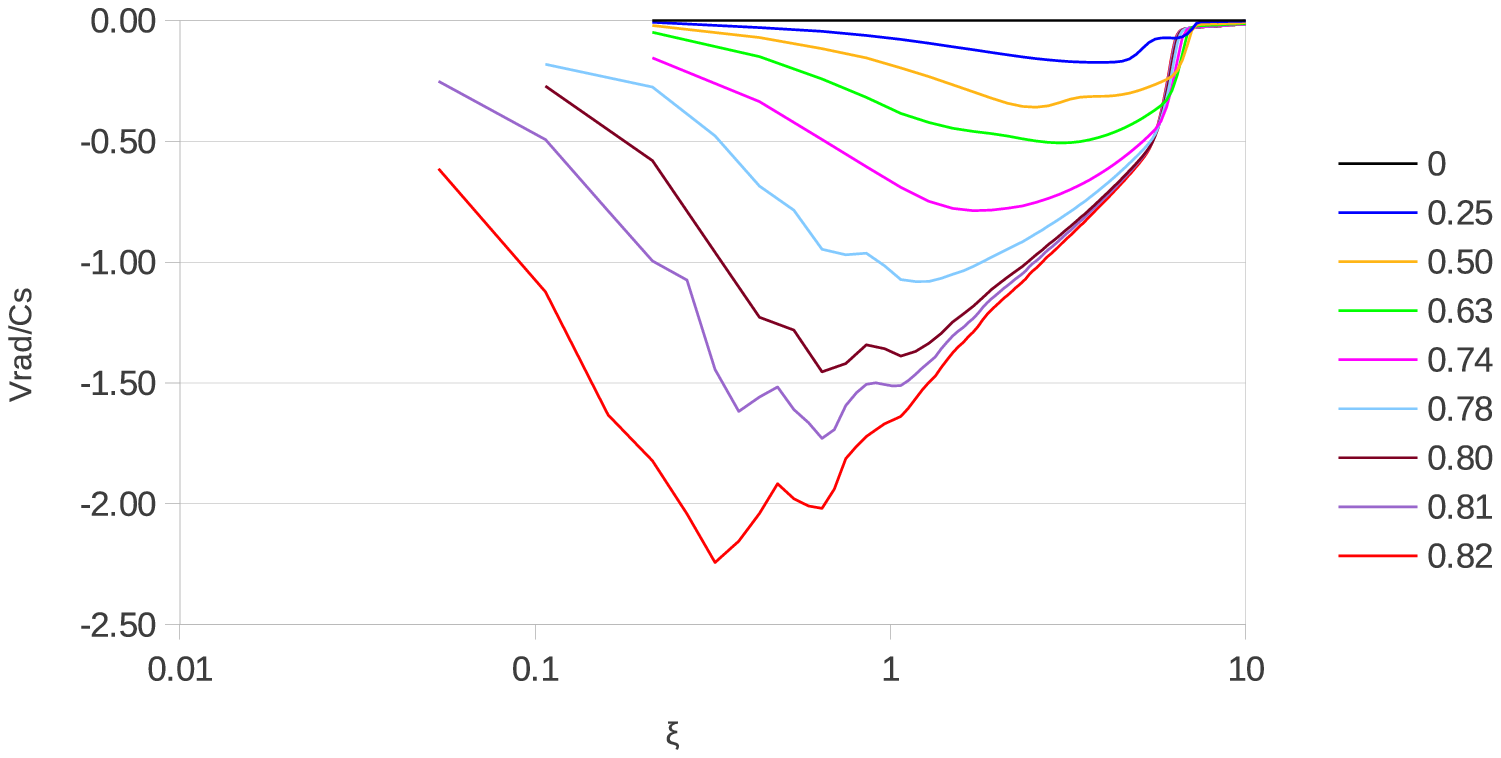}
\caption{{Density and radial velocity profiles for the ''Classic'' case, over time. The legend is time in units of the sphere's sound-crossing time, t/t$_{sc}$. As indicated by the left-hand density plot, outside-in collapse was established with the formation of a r$^{-2}$ envelope trailing a flat inner collapsing core. The right-hand radial velocity profile also matches canonical outside-in collapse, marked by an inward radial flow that began at larger radii and moved inward with time. The peak velocity became marginally supersonic, approaching $\sim$2.2 $C_{s}$ as expected (see text). The center sink particle formed by the last time state plotted in all of the figures. Here, $t_{sink}/t_{sc}$=0.82.}}
\label{BP_case}
\end{figure}

The collapse began in the outer regions of the sphere and propagated inward with time modifying the density and radial velocity profiles (Fig. \ref{BP_case}).  Note that while the times plotted in Figure \ref{BP_case} are given in units of $t_{sc}$, they are chosen relative to the formation of a sink particle. That is, the last time state plotted in Figure \ref{BP_case} is equal to the moment at which a sink formed for that particular run. Figure \ref{BP_case} shows the characteristic changes in density for BE collapse, i.e. the monotonic increase in the density of the central flat core region.  We also saw the flat core decrease in radial extent over the simulation.  Outside of the core, an infalling envelope with $r^{-2}$ profile was formed.  By t/$t_{sc}$ = 0.82 a sink particle was created and the peak inward radial velocity had become marginally supersonic ($v_{rad}\approx2.2~C_{s}$), in agreement with other studies of BE collapse \citep{banerjee2004, foster1993, hunter1977}.  We note that the maximum density at sink particle formation depends solely on our choice of maximum refinement reached ($\vartriangle$x$~\varpropto$$~\lambda_{J}$$~\varpropto$$~\rho^{-1/2}$). Since each simulation had the same maximum level of refinement, the density that triggered the final refinement level (hence sink particle formation and end of simulation) was the same throughout the different runs, i.e. $n~=~1.9\times10^ 6 ~cm^{-3}$. Note this is lower than the density $n~=~10^{7} ~cm^{-3}$ at which the isothermal assumption is no longer valid \citep{banerjee2004}.

\subsection{Collapse Varying Ambient Conditions}\label{NewModels}

Having established the code's ability to recover both hydrostatic solutions and previously explored BE collapse dynamics when triggered by an arbitrary density perturbation, we now present results for collapse under more self-consistent conditions.  In these simulations we adjusted the setup of \S~\ref{classic} model in two ways. First, we matched the density profile across the BE sphere-ambient boundary (while maintaining pressure equilibrium). Second, we allowed the simulation to progress without added perturbations. In addition to this ''Matched'' case, we also present unperturbed evolution under 3 intermediate ambient densities evenly log-spaced between the Sparse and Matched density profiles.

\subsubsection{Matched Ambient Medium}
   
\begin{figure}[htbp]
\centering
\plottwo{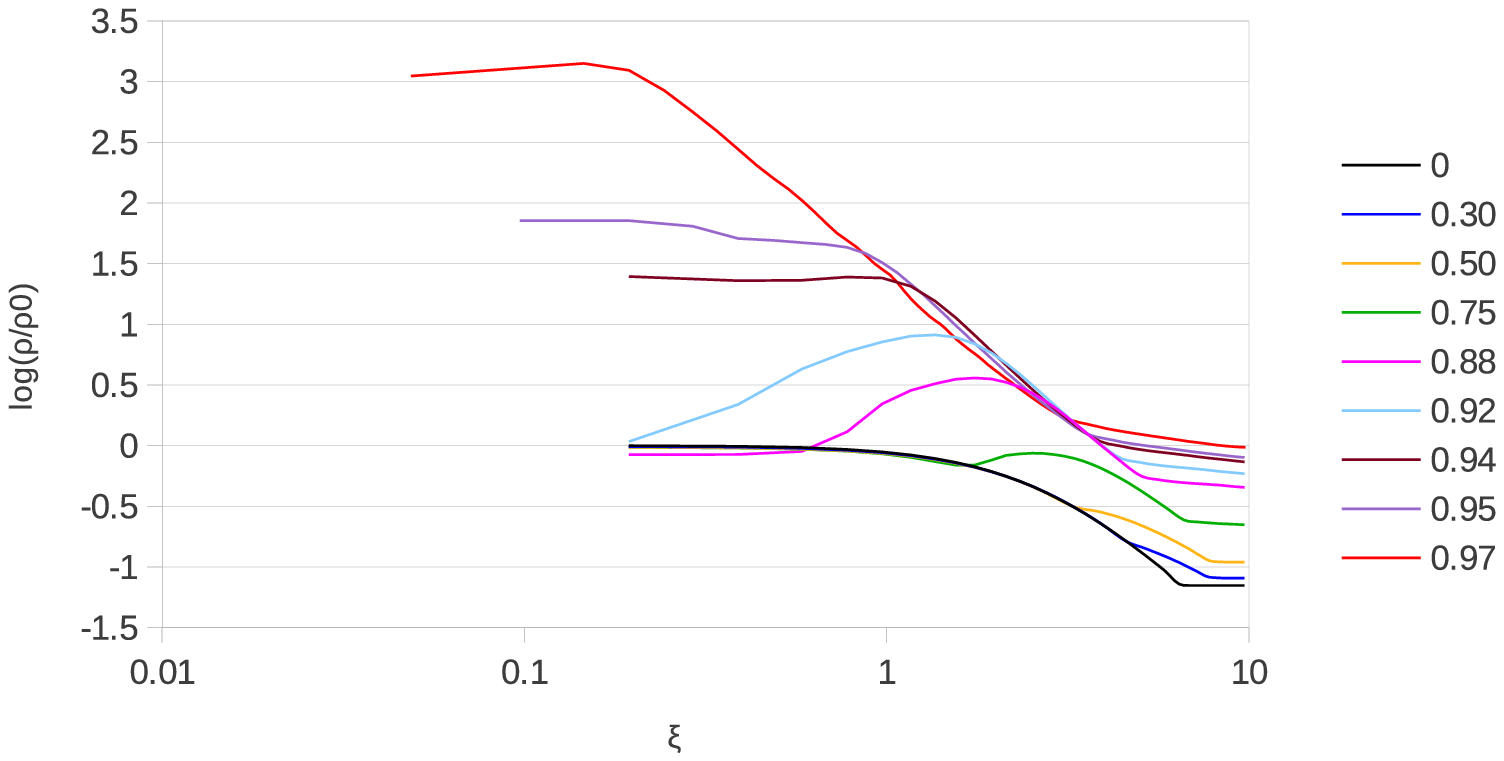}{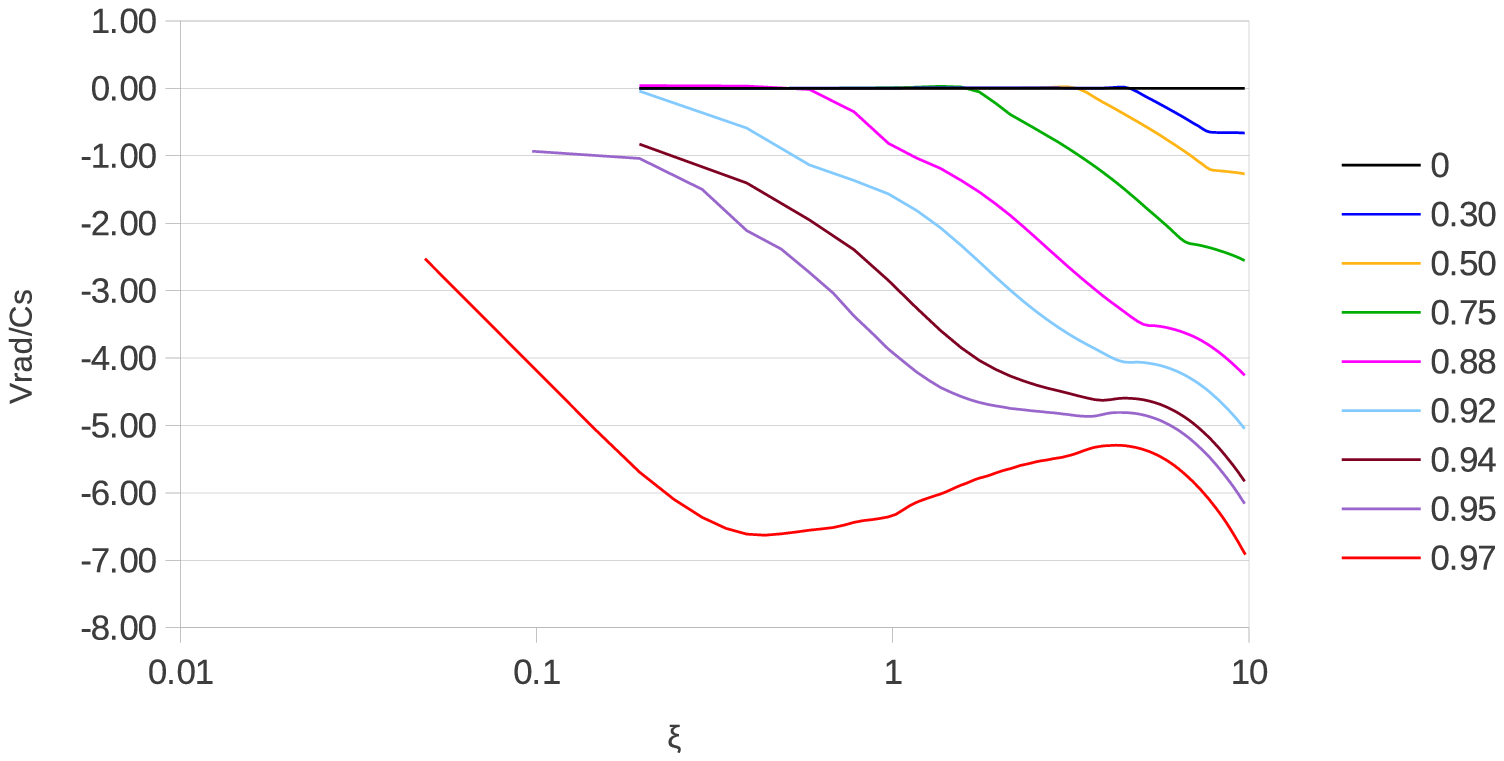}
\caption{{Density and radial velocity profiles for the Matched run, $\rho~(R_{BE})=\rho_{amb}$, over time. The legend is time in units of the sphere's sound-crossing time, t/t$_{sc}$. The density profile (left) shows material building up on the BE sphere.  Accumulated mass triggers a compression wave and collapse. By t/$t_{sc}$=0.94 the dynamics of the Classic model are recovered. The radial velocity plot is shown at right. As in the density plot, late time states show velocity profiles resembling the Classic collapse dynamics within $\xi \approx$3 by t/$t_{sc}$=0.94.}}
\label{Fig_Matched}
\end{figure}

When the density in the ambient medium is equal to that at the outer radius of the BE sphere we see qualitatively different behavior leading to collapse.  As can be seen in the density plot (Fig. \ref{Fig_Matched}, left), the Matched solution is marked by an initial infall of material from the ambient medium onto the BE sphere.  This is to be expected for two reasons.  First the sphere represents a gravitating source for the ambient material.  In addition, and more importantly, the higher density ambient material leads to a shorter collapse time based on its own self-gravity. Given that this run has the greatest ambient density it also has the shortest ambient free-fall time,

\begin{equation}\label{freefall}
t_{ff} \approx \frac{1}{4} (\frac{3 \pi}{2G \rho})^{1/2}
\end{equation}

\noindent when approximated by a uniform sphere of equal spatial extent. Since this timescale is on the order of the simulation time $t_{sim}$, and is much less than the corresponding ambient sound crossing time (see Table \ref{tab_freefalls}), it is clear that the ambient medium experiences its own homologous collapse. This can be seen as the continual increase in density of material on the sphere's outer surface.

The evolution in the Matched case is, therefore, dominated in the early stages by the infall of the ambient medium.  In Figure \ref{Fig_Matched}, we see negative velocities (infall) develop in the ambient gas once the simulation is initiated. Material initially builds up on the surface of the BE sphere and it is both the gravitational force of this material and its inward directed ram pressure that eventually overwhelms the sphere's hydrostatic equilibrium and drives it into collapse.  The explicit agent initiating collapse in this case is an inward directed compression wave induced at the BE sphere surface.

Examining the profiles for $t~ =~ 0.3 ~t_{sc}$, we see the raining down of ambient gas onto the BE sphere has increased the density at its outer edge. In particular, the inflection of the density representing the transition from the BE sphere to ambient gas has increased from $r ~=~ 1.62~pc ~ = ~R_{BE} ~(\xi ~=~ 6.5)$, to $r \approx 2 ~pc ~(\xi \approx 8 )$.  This represents a shell of new material that has been added to the BE sphere which acoustic waves must attempt to redistribute.  By $t ~=~ 0.75 ~t_{sc}$, the outer radius of the BE sphere (the location of the kink in density) has returned to its initial value and a strong peak in the density profile at $r \approx 0.6 ~pc ~(\xi \approx 2.5)$ is now apparent.  This feature marks the inward traveling compression wave. Note that as ambient material falls inward from larger radii, the density at the boundary increases, as would be expected for the collapse of a homologous sphere. Thus, the BE sphere finds itself being crushed by an inflow of ever increasing density. We note that consideration of the ambient flow on larger scales confirms it takes the form of a homologous collapse.

By $t = 0.88~t_{sc}$, the compression wave within the sphere has altered the density of what remained of the original BE density profile. By $t~ =~ 0.94 ~t_{sc}$, the dynamics {\it switches to that seen in our Classic model}. The inner flat core increases in density and shrinks in radial extent while the outer regions form an infalling envelope whose density profile does not change. A sink particle eventually formed in the inner most cell of the simulation by $t~=~0.97 ~t_{sc}$  ($14\%$ longer than in the Classic case). 

Consideration of the radial velocity profiles also demonstrates the change from the the compression wave phase to the later Classic outside-in phase. During the compression wave phase ($t/t_{sc} ~=~ 0-0.88$), inward directed fluid motions originated in the ambient gas. These negative velocity regions moved inward with the peak of the compression wave. By about $0.5 ~t_{sc}$, the flow became supersonic.  For most of the early evolution, the highest velocities are seen in the infalling ambient gas. By $t/t_{sc}~=~0.94$, however, a collapsing core within the BE sphere is established.  By the time of sink formation, the peak velocity inside the sphere corresponded with the edge of the flat core region as in the Classic case. In contrast, however, this flow has a peak mach number of $M \approx 7$, compared to the $M \approx 2.2$ of the Classic simulation.

Thus, collapse was triggered in the Matched case by the ambient gas without an applied perturbation.  To explore the limits of this ambient triggering, we now present simulations with ambient backgrounds of intermediate densities between the Matched and Sparse cases corresponding to $\rho_{amb} = \rho(R_{BE}) /3$, $\rho(R_{BE}) /10$, and $\rho(R_{BE}) /30$.

\subsubsection{Intermediate Run 1/3}

\begin{figure}[htbp]
\centering
\plottwo{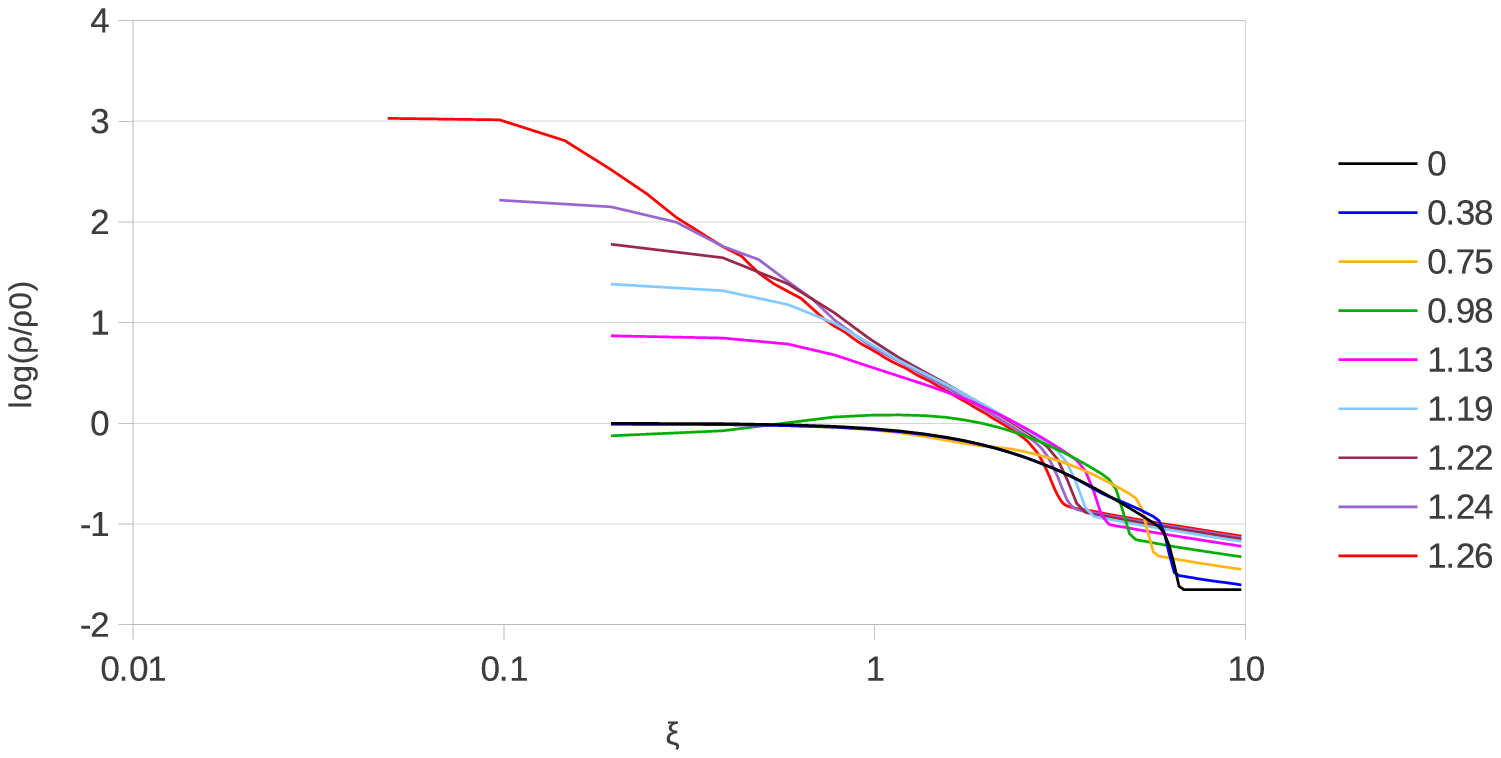}{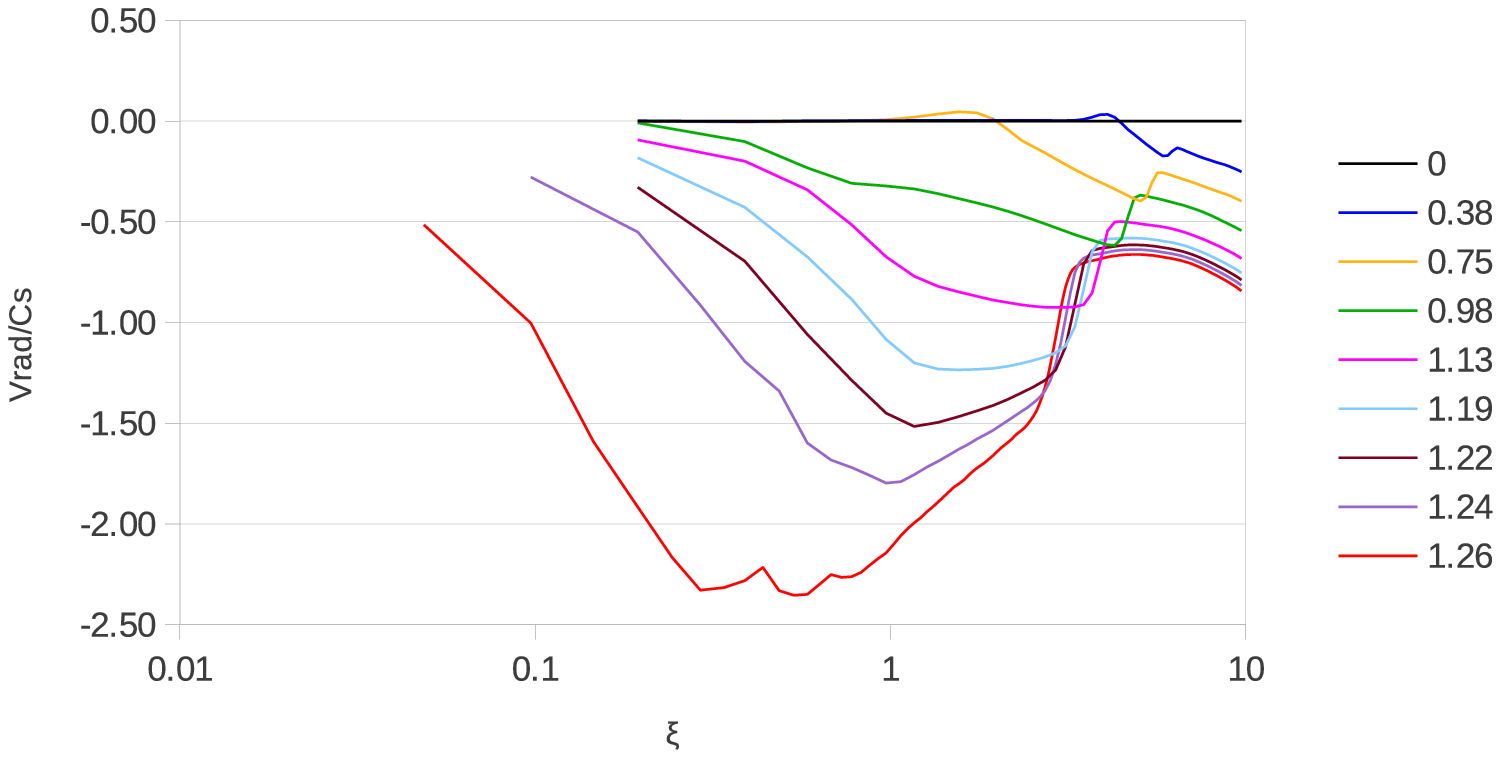}
\caption{{Density and radial velocity profiles for Run 1/3, over time. The legend is time in units of the sphere's sound-crossing time, $t/t_{sc}$. A weak density wave traveled through the BE sphere after the ambient material sufficiently collected at the BE sphere surface ($t/t_{sc} ~=~ 0 - 0.98$), as can be seen in the left-hand plot. The BE sphere was squeezed into a smaller radius with increased density, driving gravitational instability that results in collapse.  The classic outside-in collapse dynamics are achieved as evident by the shrinking flat core region and extended $r^{-2}$ envelope ($t/t_{sc}~=~1.13-1.26$). Coeval radial velocity profiles (right) show the flow to slowly become marginally supersonic and approach the Classic results inside a truncated region of the initial BE sphere.\label{w3_case}}}
\end{figure}

Evolution of the BE sphere in the densest of our Intermediate runs ($\rho_{amb}=\rho(R_{BE}) /3$) proceeds similarly to the Matched case, but on different timescales. Consideration of the density plots in Figure \ref{w3_case} shows an initial accumulation of ambient material on the BE sphere-ambient boundary (which can be tracked by the discontinuity at this edge). By $t/t_{sc}~=~0.75$, the BE sphere decreased in radial extent as well as increased in density at its outer edge (via a subsonic adjustment, evident in the corresponding radial velocity profile). As material continued to rain down on the BE sphere, its outer radius decreased and mass accumulated in its outer regions ($t/t_{sc}~=~0.98$ ). The coeval velocity profiles show the flow to be subsonic up to and including this time. Thus, we see what appears to be a subsonic compression propagating inward from the (shrinking) outer radius of the BE sphere.

Unable to incorporate the new ambient material into an approximate hydrostatic equilibrium distribution, the sphere is unable to support itself against gravitational collapse. Thus, we see collapse underway by $t/t_{sc}~=~1.13$. As in the Matched case above, the collapse resembled the Classic BE collapse in this later phase of evolution ($t/t_{sc} ~=~ 1.13-1.26$). As can be seen in the density plots, elements of the Classic collapse, such as a shrinking inner flat core trailed by an $\sim r^{-2}$ density envelope, are associated with what appears to be a smaller and denser counterpart to the original BE sphere. The velocity profiles in this phase of evolution also approach the outside-in profiles within this ''modified'' BE sphere ($\xi < 4$). 

Compared to the Matched case, there are some important points to note. First, the relative times for each phase of the evolution are different. For the Matched case, the compression phase corresponds to the early equilibration and subsequent compression wave propagation. This is the dominant phase of the Matched Case evolution, taking up a larger fraction of the total simulation time. In other words, the simulation switched to the Classic phase later in time in the Matched case than in Run 1/3. This occurred because of the stronger perturbation afforded by the ambient material in the Matched case. Second, the slower flow in Run 1/3 (only marginally supersonic) led to a longer collapse time (i.e. the time until sink particle formation) of $t_{sink}/t_{sc}=1.26$ compared to the Matched case's $t_{sink}/t_{sc}=0.97$.

It is also interesting to note that Run 1/3 was comparable to the Classic collapse in a number of ways. If we assign the beginning of the Classic collapse phase for Run 1/3 to be $t/t_{sc}~=~1.19$, we see an associated $\log (\rho / \rho_0) \approx 1.2$ and $v_{rad}/C_s \approx -1.25$. Looking at the same part of the evolution in the Classic case (now corresponding to $t/t_{sc}~=~0.78$), we see that $\log(\rho / \rho_0) \approx 1.8$ and $v_{rad}/C_s \approx -1.1$. By the end of the simulation, the peak velocities are nearly the same, $v_{rad}/C_s \approx -2.25$, although again the simulation evolved over longer time scales for this intermediate case compared to the Classic model which had $t_{sink}/t_{sc} ~=~ 0.82$. Thus, once the perturbation provided by the ambient gas was established, the flow followed the classic outside-in evolution.

\subsubsection{Intermediate Run 1/10}

\begin{figure}[htbp]
\centering
\plottwo{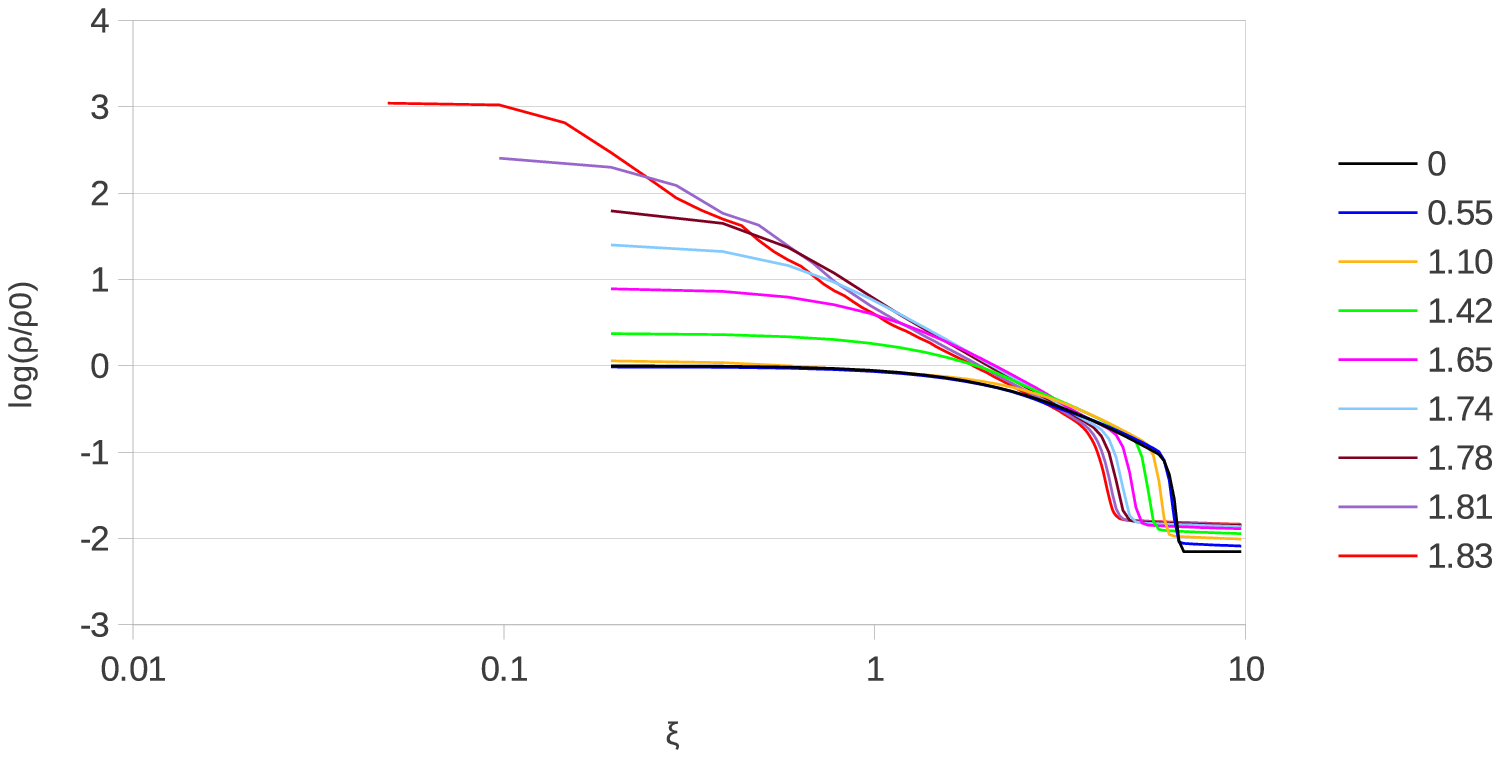}{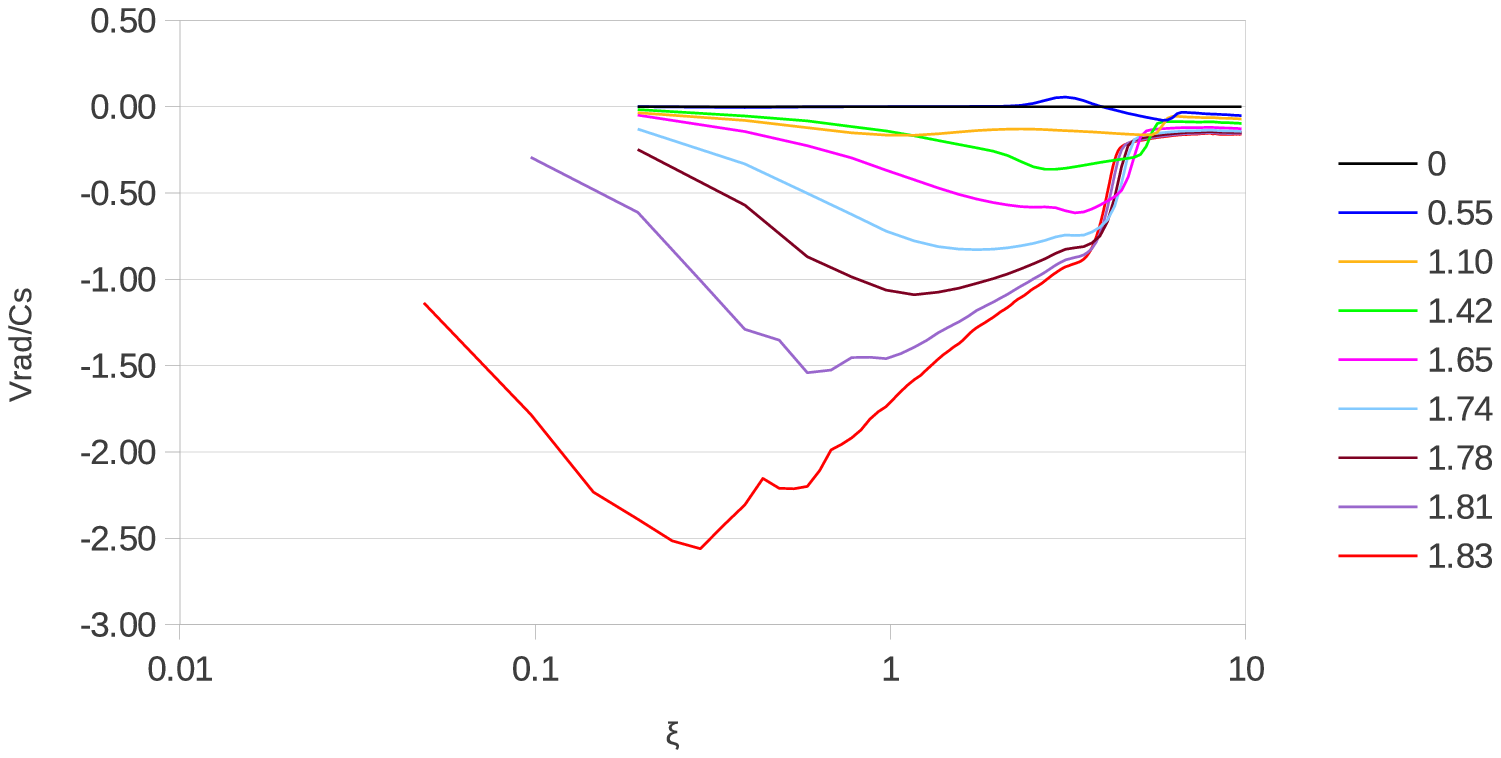}
\caption{{Density and radial velocity profiles for Run 1/10, $\rho_{amb} = \rho(R_{BE}) / 10$, over time. The legend is time in units of the sphere's sound-crossing time, $t/t_{sc}$. A light rain of ambient material accumulated on the BE sphere surface ($t/t_{sc}=0.55$) as can be seen in the density plot, left. The coeval radial velocity plot, right, shows small subsonic adjustments to this ambient flow. Amassed material squeezes the BE sphere into a smaller radial volume ($t/t_{sc}=1.10-1.42$), eventually triggering collapse. The collapse, as seen from both density and radial velocity plots, proceeds via the classic outside-in dynamics ($t/t_{sc}=1.65-1.83$).\label{w10_case} }}
\end{figure}

In a sparser ambient medium of $\rho_{amb}=\rho(R_{BE}) /10$, the evolution and the collapse of the BE sphere proceeded qualitatively differently than in the Matched and 1/3 runs. In the density plot of Figure~\ref{w10_case}, we see less accumulation of material on the outer surface of the BE sphere at $t/t_{sc}~=~0.55$. The increase in density there contributed to a slight gravitational contraction that reduced the BE sphere radius ($t/t_{sc}~=~1.10$). Further accumulation of ambient material falling onto the BE sphere after this time was negligible, yet the sphere continued to contract. As inner regions grew in density ($t/t_{sc}~ =~ 1.42$) the sphere eventually began to collapse under self-gravity when it exceeded its critical mass. The collapse then proceeded classically until sink particle formation at $t_{sink}/t_{sc}~=~1.83$. 

Consideration of the velocity profiles in Figure~\ref{w10_case} shows a subsonic adjustment of the BE sphere to material accumulating on its surface ($t/t_{sc}~=~0.55$). By $t/t_{sc}~=~1.42$, outside-in collapse is clearly underway, reaching a peak velocity close to the Classic result of $v_{rad}/C_{s} \approx -2.5$ by sink particle formation as in run 1/3.

\subsubsection{Intermediate Run 1/30}

\begin{figure}[htbp]
\centering
\plottwo{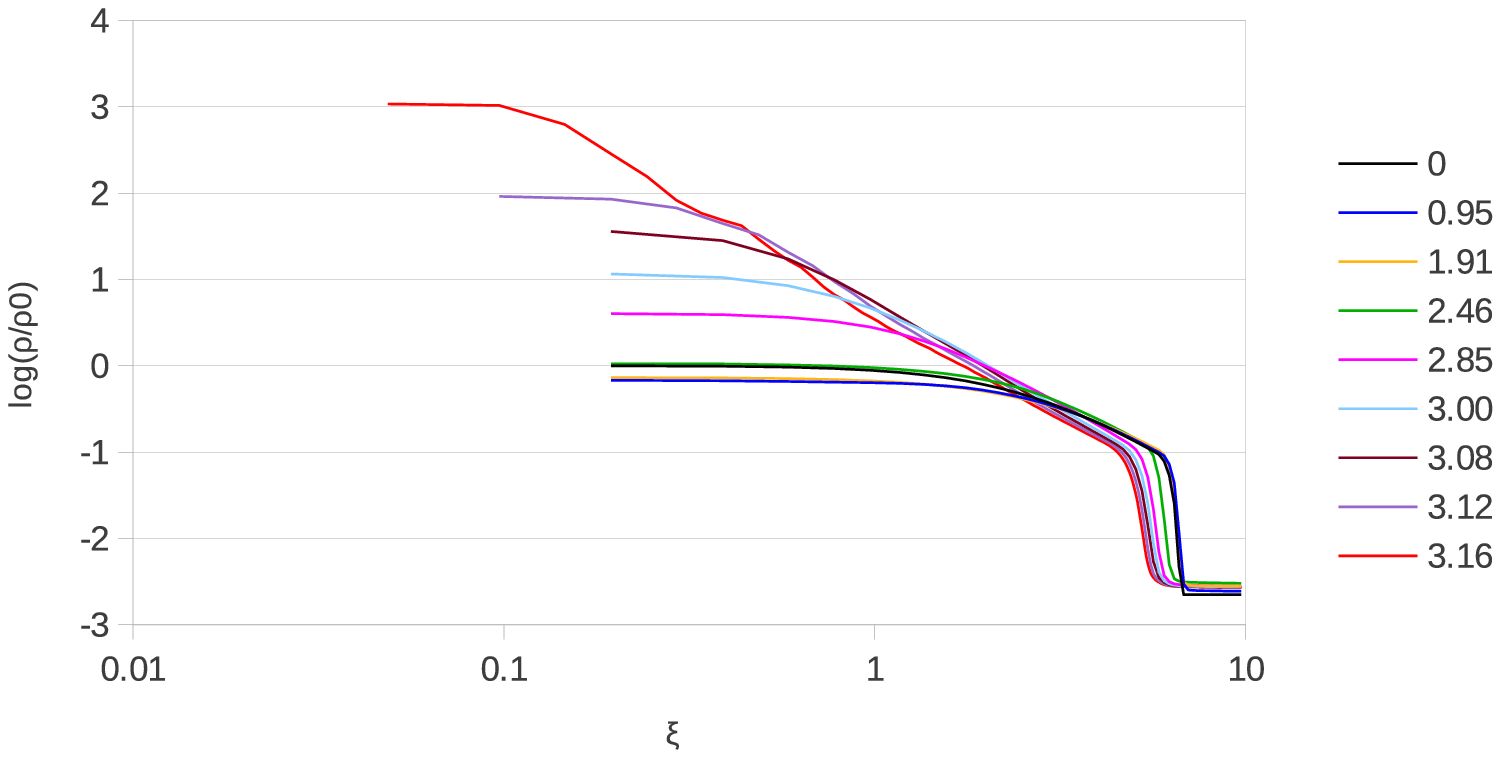}{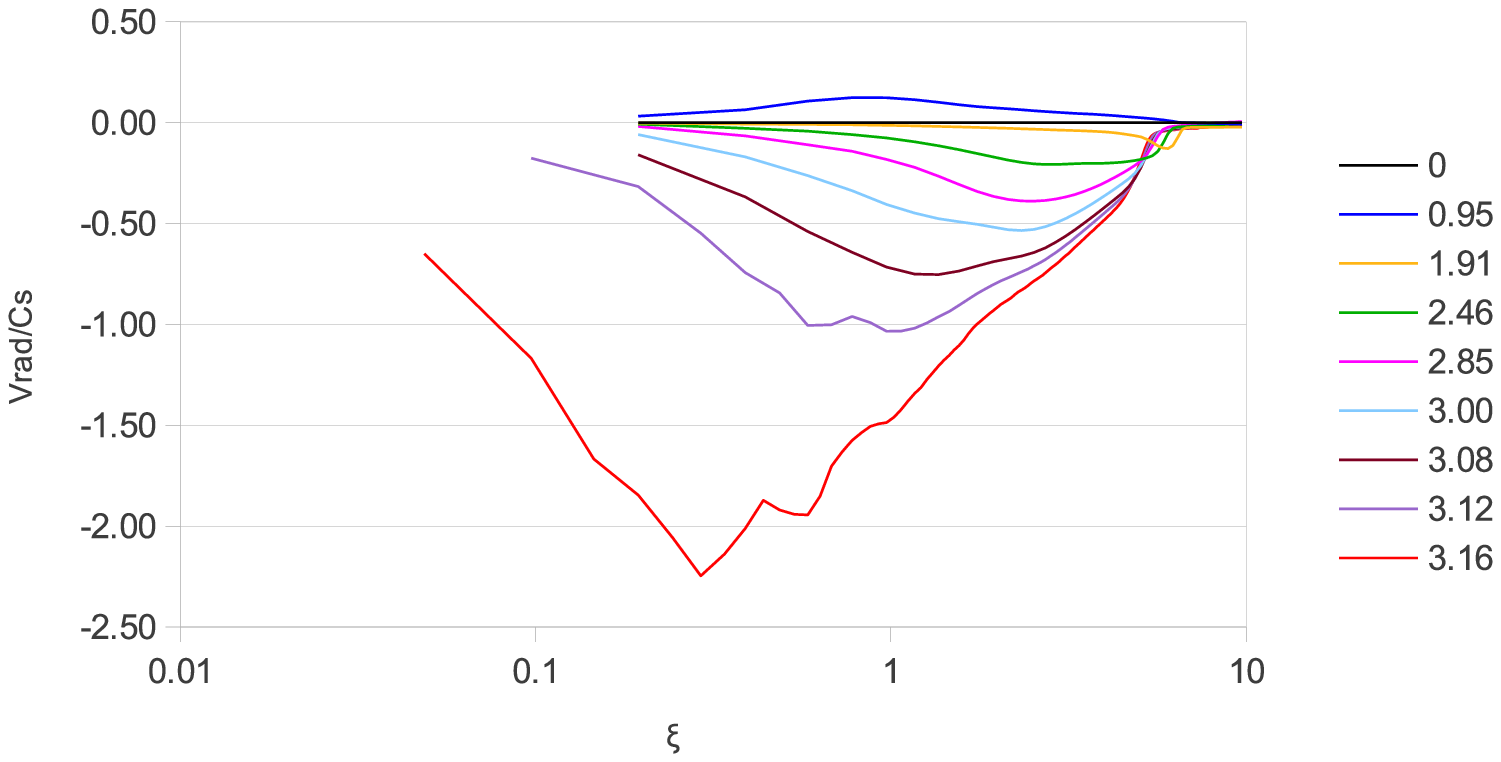}
\caption{{Density and radial velocity profiles for Run 1/30, $\rho_{amb}=\rho(R_{BE}) /30$, over time. The legend is time in units of the sphere's sound-crossing time, $t/t_{sc}$. As can be seen in the left density plot and right velocity plot, the BE sphere enters a breathing mode for time states $t/t_{sc} ~=~ 0 - 2.46$, where the interior density rises and falls and the radial velocity exhibits small subsonic oscillations. By $t/t_{sc}~=~2.86$, gravitational instability has set in and Classic outside-in collapse is underway. Collapse concludes with sink particle formation at $t_{sink}/t_{sc}~=~3.16$. \label{w30_case}}}
\end{figure}

\begin{figure}[htbp]
\epsscale{0.60}
\centering
\plotone{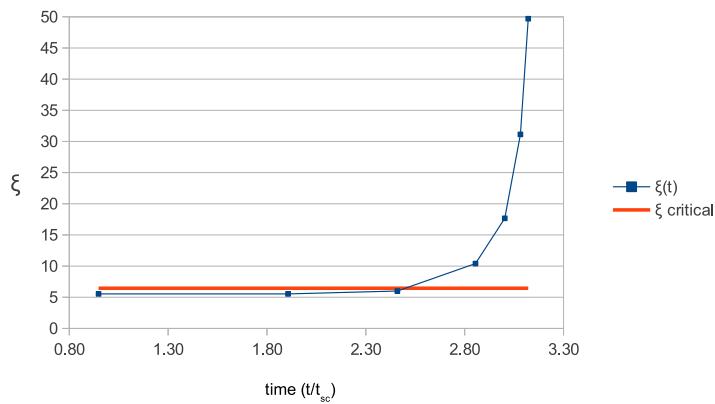}
\caption{{Variation of non-dimensional radius with time for Run 1/30. Non-dimensional radius ($\xi$) is calculated for each time shown in Figure \ref{w30_case} according to $\xi = (4\pi G \rho_0 / C_s^2)^{1/2} r$ (eqn. \ref{xi}), with $\rho_0$ and $r$ being those instantaneous values of the BE sphere for that time state. Since $\rho_0$ and $r$ change over the course of the simulation (refer Fig. \ref{w30_case}), we see a change in $\xi$ that takes the BE sphere from the sub-critical to super-critical regime by $t/t_{sc}~=~2.85$. \label{xi_crit}}}
\end{figure}

Unlike the compression wave collapse cases, but similar to the 1/10 case, the BE sphere enters a breathing mode over the first sound crossing time as evident by the outward motions of the $t/t_{sc}~=~0.95$ radial velocity curve in Figure \ref{w30_case}. During this oscillation, we see the interior density ($\xi<4$) decrease without a concomitant decrease in the BE sphere radius (location of the kink in density), indicating the breathing mode initially vacates mass from the inner regions of the BE sphere. This mass is small relative to the mass of the sphere (recall, $M_{BE}\approx 150 M_\sun$), only about 0.4 $M_\sun$.  Mass from the sphere is redistributed smoothing out local inhomogeneities. The ambient material then re-equilibrates to a rough HSE.  The re-equilibration, however, leaves the the BE sphere in the unstable regime with the non-dimensional radius $\xi > \xi_{crit}$ (Fig. \ref{xi_crit}).  Thus, even numerical noise will be enough to trigger the collapse of the re-configured sphere.  For $t>2.46$, we see the sphere experience the Classic outside-in collapse with supersonic inflows initiated by $t/t_{sc}=2.85$.

\section{Discussion}

We have found that the collapse of the Bonnor-Ebert sphere strongly depends on the nature of its encompassing medium. For those media with densities large enough to collapse on timescales comparable to the BE sphere, strong compression waves can be induced on the surface of the sphere, mediating its own collapse. For densities low enough, the direct role of the ambient medium is diminished and instead a slow increase in pressure at the sphere's edge (as well as an accumulation of ambient mass) can be the trigger of collapse. Depending on the density of the ambient material, early periods of the collapse follow one of two discrete modes discussed below. By late times, however, the collapse begins to follow the classic outside-in profile in all cases. 

One collapse trigger was a compression wave which was most strongly present in the Matched case. To understand this consider that for a sufficiently large ambient medium, the collapse time of the ambient can be approximated by the free-fall time ($t_{ff}$) of a uniform sphere (eqn. \ref{freefall}). For the Matched ambient medium this time is on the order of the BE sphere's sound crossing time, $t_{sc}$ ($t_{ff}/t_{sc} \approx 4.3~Myr/3.5~Myr \approx ~1$), indicating that the time needed for the sphere to adjust thermally via sound waves to gravitational instability induced at its boundary is comparable to the collapse time of the ambient gas. In addition, a sufficiently dense ambient medium (i.e. the Matched case) will be Jeans unstable . That is, the sound crossing time {\it of the ambient medium} is much greater than $t_{ff}$ ($t_{sc,amb}/t_{ff} \approx 130$, in our Matched case). If the ambient medium is gravitationally unstable on timescales comparable to the evolution of the BE sphere we can expect strongly driven inward motions onto the BE sphere to drive compression and eventual collapse of the sphere.
 
In contrast, when compression waves do not form we found collapse was triggered by the accumulation of infalling ambient mass. The sphere was slowly compressed ultimately changing configuration from marginally stable to unstable. To see this effect we measured the nondimensional radius $\xi$ of evolving BE spheres in our simulations. As the ambient gas both exerted increasing thermal and ram pressure on the BE sphere, $\xi$ was driven above the critical value of $\xi=6.5$ (Fig. \ref{xi_crit}) and collapse ensued.

If the density contrast between the BE sphere's outer radius and the ambient medium was made high enough, however, we found collapse was not triggered. Instead the BE spheres remained dynamically stable (as in the Sparse case, $\eta = \rho_{BE}/\rho_{amb}  \sim 100$) over many BE sphere crossing times. This indicates a critical value of the ambient density, defining a turnover between too heavy to sustain the the BE sphere and too light to induce collapse. We ran additional simulations between $\eta=30$ and $\eta=100$ to find a critical value for triggering collapse. Within the simulation time $t_{sim}$ (taken to be the $t_{sim}$ used in the $\eta = 100$ stability case), we found that $\eta = 45$ did collapse, but $\eta = 50$ did not. Running the $\eta = 50$ case out longer showed it did eventually collapse. This indicates that the turn-over occurs due to slow gravitational settling of the ambient material onto the BE sphere. Within some time range, the amount of material that has fallen onto the sphere may or may not be sufficient to induce collapse. 

To understand this behavior and to get an estimate for where the turn-over between stable and unstable configurations should occur, we model the ambient gas as a slowly settling atmosphere responding to the gravity of the BE sphere. We imagine at $t=0$ the ambient medium begins to fall in toward the surface of the BE sphere. The material that reaches the surface is heated by compression and then expands to create an atmosphere in hydrostatic equilibrium of scale height $h$. For an isothermal, plane parallel atmosphere, the scale height is given by,

\begin{equation}
h = \frac{K_B T}{m_H g}
\end{equation}

\noindent where $K_B$ is the Boltzmann constant, $T$ is the temperature, $m_H$ is the mass of a hydrogen atom, and $g$ is the local acceleration due to gravity. Assuming g is constant, we can approximate g as,

\begin{equation}
g = \frac{G M_{BE}}{R_{BE}^2}
\end{equation}

\noindent where $M_{BE}$ and $R_{BE}$ are the initial mass and radius of the BE sphere. This atmosphere induces an increased pressure on the surface of the BE sphere, which grows with time as the amount of mass that has fallen into the atmosphere increases. The pressure at the surface can then be approximated as,

\begin{equation}
P_{atm}(t) = F/A = \frac{G M_{BE} M_{atm}(t)}{4 \pi R_{BE}^2 (R_{BE} + h)^2}.
\end{equation}

\noindent To get an approximation for the mass of the atmosphere, $M_{atm}(t)$, we consider the collection of mass parcels that have had enough time to reach the BE sphere via freefall. Considering each parcel of gas in the ambient to only be gravitationally attracted by the BE sphere, the freefall time for each parcel is approximately,

\begin{equation}
t_{ff} = \frac{\pi}{2}\frac{(r)^{3/2}}{\sqrt{2 G M_{BE}}}.
\end{equation}

\noindent Inverting this equation for the freefall radius gives,

\begin{equation}
r_{ff} = (\frac{2\sqrt{2 G M_{BE}}}{\pi}t)^{2/3}.
\end{equation}

\noindent The mass within this freefall radius is then given by,

\begin{equation}
M_{atm}(t) = \frac{4}{3} \pi r_{ff}^3 \rho_{amb}
\end{equation}

\noindent where we take the ambient density $\rho_{amb}$ to be its initial value. Combining this with the pressure expression above yields,

\begin{equation}
P_{atm}(t) = \frac{8}{3 \pi^2} (\frac{G M_{BE}}{R_{BE}})^2 \frac{\rho_{amb}}{(R_{BE} + h)^2} t^2.
\end{equation} 

\noindent We used this expression to calculate the pressure perturbation induced by atmospheric settling at the BE sphere surface over time. We were only concerned with those ambient media that were Jeans stable, as determined by comparing the Jeans length of the ambient environments to the box size. This is because the turn-over occurs in this limit. 

\begin{figure}[htbp]
\centering
\plotone{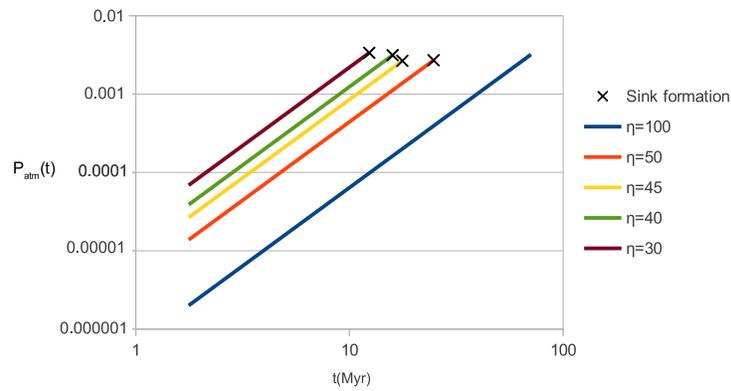}
\caption{{Pressure perturbation induced at the surface of the sphere as a function of time as derived from the atmospheric settling model. The y-axis is given in scaled units, where P$_{atm}$(ba)=P$_{atm}$(t)*Pscale and Pscale = 2.77e-13 ba. For the sparse cases $\eta=30-50$, there is a ''threshold'' value of this perturbation that is reached by sink particle formation. The $\eta=100$ case did not form a sink particle by the time this perturbation pressure should have been reached, but for this case the approximation would be expected to break down (see text). \label{pressperturb}}}
\end{figure}

Figure \ref{pressperturb} shows that our atmospheric settling model predicts a threshold value for the pressure perturbation, as indicated by the magnitude of the perturbation being nearly equal at sink particle formation in the different simulations. The model also predicts that as the ambient density is decreased ($\eta$ increased), the time it takes to reach this threshold grows. This allows a prediction of when collapse might be triggered by calculating when the pressure perturbation meets this critical value. Thus, this simple model appears to capture the threshold in $\eta$ for which configurations should collapse. The model would break down obviously when the scale-height exceeds the box size, as is the case for $\eta=100$, which remained dynamically stable beyond the time predicted. 

It should again be emphasized that the work presented here was an idealization of the system we studied. This is because of the placement of the Bonnor-Ebert sphere at the center of mass position of concentric, spherically symmetric ambient environments. This assumption is convenient, but it is of course a special case that makes the environment in the Matched case ''maximally crushing''. If the BE sphere's position did not coincide with the center of mass, or if it were embedded in a cylindrical filament instead of a sphere, the evolution might be more representative, and  less dominated by global infall. 

On the other hand, we would not expect the addition of rotation and cooling to alter the results significantly. \cite{banerjee2004} looked at the effects of rotation and cooling on the collapse of BE spheres. They showed that with these effects, the supercritical BE sphere that we modeled here (their case ‘'A1'’) collapsed just as it would have had it been purely isothermal and non-rotating, within the isothermal regime. Thus, we expect no changes in the results if we were to add cooling and rotation. Further, recall that the main difference the ambient afforded was a change in the way collapse was {\it initiated}. By the end of the simulation (and again well within the isothermal regime), density profiles resumed characteristic shapes. Thus, if we were to include cooling and rotation,  and follow the simulation further (i.e. into the {\it non}-isothermal regime), we would expect the results of our simulations to be very similar to \cite{banerjee2004}, but perhaps with slightly different timescales and morphologies. Deviations might in part be attributed to the perturbed velocity structure the early compression wave afforded the collapse.

Despite the models' simplicities, connections between our simulations and the astrophysical environment can be drawn. The Matched case bears a resemblance to the cores described by \cite{teixeira2005} in the Lupus complex, where dense cores within the star-forming Lupus 3 molecular cloud have a gradual decrease in density moving outward from the core center. In contrast, the mis-matched cases are more like static gas with a phase change from core to environment, such as cold, dense neutral globules embedded in hot, rarefied, ionized HII regions.

\section{Conclusions}

In this paper we have explored the role of the ambient medium in triggering the collapse of marginally stable Bonner-Ebert spheres.  We first confirmed that our code could both reproduce long-lived stable BE spheres and recover previously seen outside-in collapse modes driven by an arbitrarily imposed perturbation.  We then carried out a series of simulations of increasing density contrast between the BE sphere edge and the constant density ambient  medium surrounding the sphere.  For our Matched case, $\eta = \rho(R_{BE})/\rho_{amb} = 1$, we found the ambient medium was itself Jeans unstable.  Its infall drove a strong compression wave which then triggered the classic outside-in collapse of the BE sphere.  This compression wave-triggered collapse mode held for density contrasts up to $\eta = 10$.  In contrast, for $\eta > 10$, a form of gravitational settling led to an increased pressure on the BE sphere surface.  The higher pressure forced the BE sphere to adjust such that it eventually switched from being marginally stable to unstable and then followed an outside-in collapse.  For low enough ambient densities (given our computational box size) the BE spheres remained stable for the duration of the simulations.

Given there is a cut-off to the induced dynamical instability when the ambient density becomes low enough, our results indicate that BE sphere type solutions for isolated starless cores/clumps require high values of the density contrast to be long-lived.  Lower density contrasts imply the clouds will be not be stable structures.  Instead they will most likely be part of a region in which collapse is ongoing both locally and globally. Given the rapid crushing of the BE spheres by the infall of highly Jeans unstable environments, only high $\eta$ solutions may be models for long-lived cores (with properties like a BE sphere). 

While we found a distinction between the different triggers of collapse there was little difference in the late phase of evolution (i.e. after collapse had been initiated). The systems that went unstable always eventually followed outside-in dynamics that include the development of a collapsing flat-topped core surrounded by a $r^{-2}$ envelope with supersonic radial inflow that tracks the core-envelope boundary. Thus, we conclude that the well known features of BE sphere collapse are robust after collapse is initiated. Furthermore, these results provide support for the WS framework (\cite{whitworth1985}) in matching the predicted roles of compression waves with the collapse of embedded BE spheres.  Our work is also compatible with the studies of \cite{hennebelle2003}, who looked at the effect of differentially increasing the external pressure on marginally stable BE spheres. They found the trigger of collapse to be strong compression waves when the external pressure was increased rapidly.

Our work emphasizes the likely unitary nature of star forming environments such that separations between core and inter-core medium may be largely a matter of definition only.  For our simulations, it appears that unless cores are formed with very large density contrasts relative to their surroundings, ambient gas will play an important and ongoing role in the core's dynamics.  This is in line with models that attempt to consider more continuous distributions of variables between core and environment \citep{myers2008}. 

\acknowledgments 

We are grateful for the support provided by the Space Telescope Science Institute through grants HST-AR-12128 and HST-AR-12832, the Department of Energy through grant numbers DE-SC0001063 and R17081, the National Science Foundation for the Extreme Science and Engineering Discovery Environment (XSEDE) through grant number OCI-1053575 and for grant number AST-1109285. We would also like to thank the University of Rochester's Laboratory for Laser Energetics for funds received by the Horton Fellowship and the University's Center for Integrated Research Computing for providing the supercomputer resources that supported this work. 

\appendix

\section{Appendix}

\subsection{Self-gravity in Astrobear2.0}\label{poissonsolver}

The Euler equations with self-gravity are given by,

\begin{equation}\label{cont2}
\frac{\partial \rho}{\partial t} + \pmb{\nabla} \pmb{\cdot} (\rho \textbf{v}) = 0
\end{equation}

\begin{equation}\label{mom2}
\frac{\partial \rho \textbf{v}}{\partial t} + \pmb{\nabla \cdot} (\rho \textbf{v} \textbf{v} + P\pmb{I}) = - \rho \pmb{\nabla} \phi
\end{equation}

\begin{equation}\label{energy2}
\frac{\partial E}{\partial t} + \pmb{\nabla \cdot} [\pmb{v}(E+P)] = -\rho\pmb{v\cdot\nabla}\phi
\end{equation}

\noindent where $\rho$ is the mass density, P is the pressure, $\vec{v}$ is the vector of velocities, \textit{\textbf{I}} is the unit tensor, $\phi$ is the gravitational potential satisfied by Poisson's equation,

\begin{equation}\label{grav2}
\nabla^2 \phi = 4 \pi G (\rho - \bar{\rho})
\end{equation} 

\noindent and E is the total energy given by,

\begin{equation}
E = \frac{1}{2}\rho v^2 + \epsilon \rho
\end{equation}

\noindent where $\bar{\rho}$ is the mean mass density and $\epsilon$ is the specific internal energy. For ideal gas, the caloric equation of state gives $\epsilon = P/[(\gamma - 1)\rho]$, where $\gamma$ is the ratio of specific heats. Subtracting off the mean density ($\bar{\rho}$) in equation (\ref{grav2}) is necessary when solving for the potential on a periodic grid, as it provides an additional condition to close the system of equations. For non-periodic grids, the code sets $\bar{\rho} = 0$ for this condition. 

Astrobear can solve the self-gravitating fluid equations (\ref{cont2}-\ref{energy2}) as is using a Strang split method + a corner transport upwind (CTU) scheme, or can solve the equations with momentum conservation. Using a Strang split method, the homogeneous versions of the Euler equations is solved during the hydrodynamic steps and the source term equations, $d\pmb{u}/dt=\pmb{s}(\pmb{u})$ are solved during the source steps, where $\pmb{u}$ is the vector of fluid variables and $\pmb{s}$ is the source-term vector given by the components on the right hand side of equations (\ref{cont2}-\ref{energy2}). Here, a source step of half a time step is taken on both sides of the hydrodynamic advance step to make the scheme 2nd-order accurate in time. The hydrodynamic advance steps can use any one of many available hyperbolic solvers. 

Using just a Strang splitting method, we noticed sink particles that formed under spherically symmetric conditions would occasionally wander across the grid, even though the net forces on the sink should have been zero. To correct this, we enforced strict momentum conservation across grids. This eliminated the erroneous kicks to the sink particles. To implement a momentum conserving scheme, the fluid equations must be expressed in conservative form. That is, in terms of a tensor flux function \textit{\textbf{F}} for which the following relationship holds,

\begin{equation}
\frac{\partial \pmb{u}}{ \partial t} = - \pmb{\nabla \cdot F}
\end {equation}

\noindent For momentum conservation, this flux function \textit{\textbf {F}} follows from enforcing the right hand side of the momentum equation (\ref{mom2}) to be equal to the negative divergence of \textit{\textbf {F}},

\begin{equation}
\rho \pmb{\nabla} \phi = \pmb{\nabla \cdot F}.
\end{equation}

\noindent Using Poisson's equation (\ref{grav2}) to substitute for $\rho$ gives,

\begin{equation}
\pmb{\nabla \cdot F} = (\frac{\nabla ^2 \phi}{4 \pi G} + \bar{\rho}) \pmb{\nabla} \phi
\end{equation}

\noindent which in 1D is equivalent to,

\begin{equation}
\pmb{\nabla \cdot F} = \pmb{\nabla \cdot} [\frac{1}{2}\frac{(\pmb{\nabla} \phi)^2}{4 \pi G} + \bar{\rho} \phi \pmb{I}]
\end{equation}

\noindent where we can identify the equivalent momentum flux tensor as 

\begin{equation}
\pmb{F} = \frac{1}{2}\frac{(\pmb{\nabla} \phi)^2}{4 \pi G} + \bar{\rho} \phi \pmb{I}.
\end{equation}

\noindent A similar analysis gives the form of the flux function tensor for general dimension as,

\begin{equation}
F_{ij} = \frac{(\partial_j \phi \partial_i \phi)}{4 \pi G} - \delta^i_j (\frac{\partial_k \phi \partial_k \phi}{8 \pi G} + \bar{\rho} \phi).
\end{equation}

\noindent The {\it total} flux, including both this true flux and the Riemann flux calculated across cell boundaries, is used to update the momenta array during the CTU scheme.

For both the non-momentum conserving and momentum conserving schemes, we solve for the gravitational potential using the HYPRE library, a generalized software package that solves linear systems on massively parallelized computing systems. The stencil for the potential varies depending on the grid dimension and whether it has fixed cell spacing or not. In its simplest form, the 2nd-order accurate discretized version of equation (\ref{grav2}) in 1D gives,

\begin{equation}
\frac{(u_{i+1} + u_{i-1} - 2 u_i)}{(\triangle x)^2} =  4 \pi G (\rho_i-\bar{\rho})
\end{equation}

\noindent where $\triangle x$ is the intercell spacing. The potential is solved for independently on each level. To update the ghost cells of finer-level grids, we use $\dot{\phi}$ from the courser levels. Once the new potential is calculated, we apply a correction to the source terms giving 2nd-order accuracy in time.

\subsection{Testing self-gravity in Astrobear2.0}\label{testingpoissonsolver}

As a test for our self-gravity implementation, we checked whether the code could reproduce the correct growth rate for the Jeans instability. We briefly review the derivation for the growth rate here. The linearized, self-gravitating (1D) fluid equations are given by

\begin{equation}\label{continuity}
\frac{\partial \rho '}{\partial t} + \rho_0 \triangledown \cdot v ' = 0
\end{equation}

\begin{equation}\label{momentum}
\rho_0 \frac{\partial {v '}}{\partial t} = - C_s^2 \triangledown \rho ' - \triangledown \phi '
\end{equation}

\begin{equation}\label{gravity}
\triangledown^2 \phi ' = 4 \pi G \rho '
\end{equation}

\noindent where $C_s$ is the sound speed, $\rho$ is the mass density, $v$ is the velocity, and $\phi$ is the gravitational potential. The background density ($\rho_0$) and gravitational potential ($\phi$) is assumed constant, and the velocity is initially zero ($v_0 = 0$) (note this reproduces the classic Jeans swindle, but for our purposes it is safe to ignore this effect). The primed terms in the equations indicate the perturbation in the corresponding variable. By combining these equations, one arrives at the following wave equation in the density perturbation,

\begin{equation}\label{density_ODE}
\frac{\partial ^2 \rho '}{\partial t ^2} -  C_s^2 \triangledown^2 \rho ' = -4 \pi G \rho_0 \rho '.
\end{equation}

\noindent Looking for plane-wave solutions to this equation,

\begin{equation}\label{density_soln}
\rho ' \propto e^{ikx - \omega t}
\end{equation}

\noindent gives the dispersion relation, 

\begin{equation}\label{dispersion}
\omega^2 = C_s^2 k^2 - 4 \pi G \rho_0
\end{equation}

\noindent where $k$ is the wavenumber and $\omega$ is the frequency. From this we have the growth rate of the density perturbation ($\omega ^{-1}$). Note that only for $C_s^2 k^2 - 4\pi G \rho_0 < 0$, are the waves exponentially growing. This length scale defines the Jeans length, and underlines the classical Jeans instability. 

To find the corresponding velocity perturbation, we insert the determined density function into the continuity equation (\ref{continuity}). This gives, 

\begin{equation}\label{dispersion}
v ' \propto - \frac{\omega}{\rho_0 k}Sin(kx).
\end{equation}

\noindent The pressure perturbation is then specified using the ideal gas equation,

\begin{equation}\label{dispersion}
P \propto \frac{K_B T}{m_H}(\rho_0 + \rho ')
\end{equation}

\noindent with $P$ being the thermal pressure. 

\begin{figure}[htbp]
\centering
\plotone{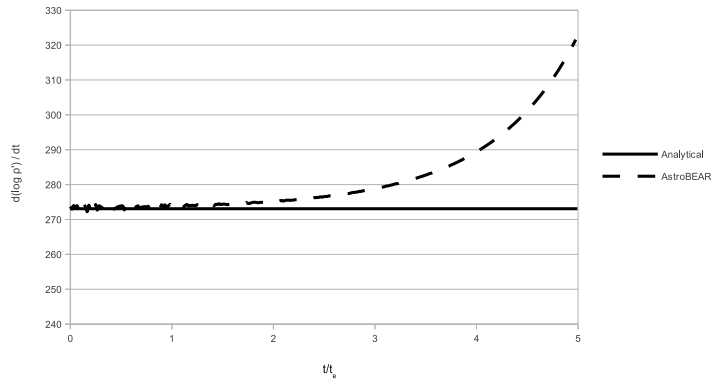}  
\caption{{Comparing Astrobear output with theory. From the text, {$\rho' \propto e^{A t}$, where $A$ is a constant, giving $\log(\rho') \propto [A log (e)] t$. The slope of this line is  $A \log e = 273$ (computational units) for the parameters of the test. Numerical data was taken from a peak in the growing instability, over time, and compared to this value. Astrobear provides a close match for the first few e-folding times $t_e$. \label{jeans_perturb}}}}
\end{figure}

We seeded a 1D grid with these perturbations, of amplitude $\delta = 0.001$, and evolved the domain over 5 e-folding times (the time it takes the perturbation to increase by a factor of 'e'), using periodic boundary conditions for both the box and the elliptic solver. The code was able to match the growth rate of the perturbation over the first couple of e-folding times ($\tau_e$), to good accuracy (Fig. \ref{jeans_perturb}). These results confirm Astrobear's ability to model self-gravity in astrophysical flows.

\bibliography{erica}

\begin{thebibliography}{27}
\expandafter\ifx\csname natexlab\endcsname\relax\def\natexlab#1{#1}\fi

\bibitem[{{Alves} {et~al.}(2001){Alves}, {Lada}, \& {Lada}}]{alves2001}
{Alves}, J.~F., {Lada}, C.~J., \& {Lada}, E.~A. 2001, nature, 409, 159

\bibitem[{{Anathpindika} \& {Di Francesco}(2013)}]{anathpindika2013}
{Anathpindika}, S., \& {Di Francesco}, J. 2013, \mnras, 430, 154

\bibitem[{{Banerjee} {et~al.}(2004){Banerjee}, {Pudritz}, \&
  {Holmes}}]{banerjee2004}
{Banerjee}, R., {Pudritz}, R.~E., \& {Holmes}, L. 2004, \mnras, 355, 248

\bibitem[{{Bonnor}(1956)}]{bonnor1956}
{Bonnor}, W. 1956, \mnras, 116, 351

\bibitem[{{Carroll-Nellenback} {et~al.}(2013){Carroll-Nellenback}, {Shroyer},
  {Frank}, \& {Ding}}]{carroll2013}
{Carroll-Nellenback}, J., {Shroyer}, B., {Frank}, A., \& {Ding}, C. 2013,
  Journal of Computational Physics, 236, 461

\bibitem[{Cunningham {et~al.}(2009)Cunningham, Frank, Varni\`{e}re, Mitran, \&
  Jones}]{andy2009}
Cunningham, A.~J., Frank, A., Varni\`{e}re, P., Mitran, S., \& Jones, T.~W.
  2009, \apjs, 182, 519

\bibitem[{{Ebert}(1955)}]{ebert1955}
{Ebert}, R. 1955, Z. Astrophys, 37, 217

\bibitem[{{Falgout} \& {Yang}(2002)}]{falgout2002}
{Falgout}, R., \& {Yang}, U. 2002, in Computational Science - ICCS 2002 Part
  III, of Lecture Notes in Computer Science, Vol. 2331 (Springer-Verlag),
  632--641

\bibitem[{{Federrath} {et~al.}(2010){Federrath}, {Banerjee}, {Clark}, \&
  {Klessen}}]{fedderrath2010}
{Federrath}, C., {Banerjee}, R., {Clark}, P.~C., \& {Klessen}, R.~S. 2010,
  \apj, 713, 269

\bibitem[{{Foster}(1994)}]{foster1994}
{Foster}, P. 1994, in Proceedings of the 4th Haystack Observatory Conference,
  Vol.~65, 105

\bibitem[{{Foster} \& {Chevalier}(1993)}]{foster1993}
{Foster}, P.~N., \& {Chevalier}, R.~A. 1993, \apj, 416, 303

\bibitem[{{Hennebelle} {et~al.}(2003){Hennebelle}, {Whitworth}, {Gladwin}, \&
  {Andre}}]{hennebelle2003}
{Hennebelle}, A.~P., {Whitworth}, A.~P., {Gladwin}, P.~P., \& {Andre}, P. 2003,
  \mnras, 340, 870

\bibitem[{{Hunter}(1977)}]{hunter1977}
{Hunter}, C. 1977, \apj, 218, 834

\bibitem[{{Keto} \& {Caselli}(2010)}]{keto2010}
{Keto}, E., \& {Caselli}, P. 2010, \mnras, 402, 1625

\bibitem[{{Larson}(1969)}]{larson1969}
{Larson}, R. 1969, \mnras, 145, 271

\bibitem[{{Liu}(1996)}]{liu1996}
{Liu}, F.~K. 1996, \mnras, 281, 1197

\bibitem[{{Myers}(2008)}]{myers2008}
{Myers}, P.~C. 2008, \apj, 687, 340

\bibitem[{{Ogino} {et~al.}(1999){Ogino}, {Tomisaka}, \& {Nakamura}}]{ogino1999}
{Ogino}, S., {Tomisaka}, K., \& {Nakamura}, F. 1999, \pasj, 51, 637

\bibitem[{{Penston}(1969)}]{penston1969}
{Penston}, M. 1969, \mnras, 144, 425

\bibitem[{{Shu}(1977)}]{shu1977}
{Shu}, F.~H. 1977, \apj, 214, 488

\bibitem[{{Shu} \& {Adams}(1987)}]{adams1987}
{Shu}, F.~H., \& {Adams}, F. 1987, in Circumstellar matter; Proceedings of the
  IAU Symposium, Heidelberg, 7--22

\bibitem[{{Shu} {et~al.}(1987){Shu}, {Lizano}, \& {Adams}}]{shu1987}
{Shu}, F.~H., {Lizano}, S., \& {Adams}, F. 1987, in {Star forming regions;
  Proceedings of the Symposium, Tokyo, Japan}, 417--433

\bibitem[{Spitzer(1968)}]{spitzer1968}
Spitzer, L. 1968, {Nebulae and Interstellar Matter} (The University of Chicago
  Press), 44

\bibitem[{{Teixeira} {et~al.}(2005){Teixeira}, {Lada}, \&
  {Alves}}]{teixeira2005}
{Teixeira}, P., {Lada}, C., \& {Alves}, J. 2005, \apj, 629, 276

\bibitem[{{Truelove} {et~al.}(1997){Truelove}, {Klein}, {McKee}, {Holliman},
  {Howell}, \& {Greenough}}]{truelove1997}
{Truelove}, J., {Klein}, R.~L., {McKee}, C.~F., {Holliman}, J.~H., {Howell},
  L.~H., \& {Greenough}, J.~A. 1997, \apjl, 489

\bibitem[{Ward-Thompson {et~al.}(2007)Ward-Thompson, Andre, Crutcher,
  Johnstone, Onishi, \& Wilson}]{ward-thompson2007}
Ward-Thompson, D., Andre, D., Crutcher, P., Johnstone, D., Onishi, T., \&
  Wilson, C. 2007, {Protostars and Planets V} (Tucson: Univ Arizona Press), 33

\bibitem[{{Whitworth} \& {Summers}(1985)}]{whitworth1985}
{Whitworth}, A., \& {Summers}, D. 1985, \mnras, 214, 1

\end{thebibliography}

\clearpage

\clearpage

\begin{table}[htbp]\centering\small
\caption{Description of the 9 simulations of various background densities.}
\begin{tabular}{ccc}
\hline
Run & Density of Ambient  ($\rho(R_{BE})$) & Perturbation? \\
\hline
Sparse  &    1/100  &  No        \\
Classic &    1/100    &   YES \\
Matched  &    1   & No      \\
1/3  &    1/3    & No       \\
1/10 &  1/10    &    No   \\
1/30 &  1/30   &   No       \\
1/40 & 1/40  & No      \\
1/45 & 1/45     & No      \\
1/50 & 1/50     & No      \\
\hline
\end{tabular}
\tablecomments{The second column gives the ambient density of the simulation in terms of the density at the BE sphere's initial outer radius, $R_{BE}$, and the third column states whether a perturbation was applied to initiate collapse.\label{tab_runs}}
\end{table}

\clearpage

\begin{table}[htbp]\centering\small
\caption{Thermal and gravitational timescales for the ambient gas along with simulation times for the different runs.}
\begin{tabular}{ccccc}
\hline\hline
Run  & $t_{ff}$ & $t_{sim}$ & $t_{sc}$ \\
 & (Myr) & (Myr) & (Myr) & \\
\hline
Classic & 41.5  & 3.3 & 6.1  \\
Matched & 4.4  & 3.7 & 57.5  \\
1/3 & 7.8  & 4.9  & 33.1   \\
1/10 & 13.8  & 7  & 19.2   \\
1/30 & 24.4  &  12.2  & 10.5   \\
\hline
\end{tabular}
\tablecomments{Recall, $t_{sim}=t_{sink}$, that is, the simulations were halted once a sink particle formed. The free-fall time ($t_{ff}$) is calculated using equation (\ref{freefall}) and the ambient density. To do so is approximating the ambient gas to be spherical and the BE sphere to be negligible, a valid approximation for our purposes here. The sound crossing time ($t_{sc}$) is calculated using equation (\ref{eqn_tsc}), but now with $r=25 ~pc$ ($L/2$) and $C_s$ of the ambient.}
\label{tab_freefalls}
\end{table}

\end{document}